\documentclass[floatfix,twocolumn,prd,tightenlines,preprintnumbers,amsmath,amssymb,superscriptaddress]{revtex4-2} 
\usepackage{amsmath}
\usepackage{bm}
\usepackage{graphicx}
\usepackage{mathrsfs}
\usepackage{multirow}
\usepackage{graphicx}
\usepackage{booktabs, ctable}
\usepackage{xcolor, color, framed}
\usepackage[colorlinks, linkcolor=red,anchorcolor=green,citecolor=blue]{hyperref}

\begin{document}
\title{Evidence for Y-shaped color confinement of fully heavy baryons}

\author{Jiaxing Zhao}
\email{jzhao@itp.uni-frankfurt.de}
\affiliation{Helmholtz Research Academy Hesse for FAIR (HFHF), GSI Helmholtz Center for Heavy Ion Physics, Campus Frankfurt, 60438 Frankfurt, Germany}
\affiliation{Institut f\"ur Theoretische Physik, Johann Wolfgang Goethe-Universit\"at, Max-von-Laue-Stra{\ss}e 1, D-60438 Frankfurt am Main, Germany}
\affiliation{Department of Physics, Yantai University, Yantai 264005, China}

\author{Shuzhe Shi}
\email{shuzhe-shi@tsinghua.edu.cn}
\affiliation{Department of Physics, Tsinghua University, Beijing 100084, China}
\affiliation{State Key Laboratory of Low-Dimensional Quantum Physics, Tsinghua University, Beijing 100084, China}

\author{Pengfei Zhuang}
\email{zhuangpf@mail.tsinghua.edu.cn}
\affiliation{Department of Physics, Yantai University, Yantai 264005, China}
\affiliation{Department of Physics, Tsinghua University, Beijing 100084, China}
\affiliation{South Center for Nuclear-Science Theory, Institute of Modern Physics, Chinese Academy of Sciences, Huizhou 516000, China}

\begin{abstract}
The geometry of color confinement in baryons remains a fundamental question in quantum chromodynamics. We investigate whether heavy-quark spectroscopy can distinguish a genuine three-body Y-shaped confining interaction from a sum of pairwise $\Delta$-shaped interaction. Using a high-precision three-body Schr\"odinger solver, we perform a simultaneous analysis of experimentally measured bottomonium masses and lattice-QCD calculations of $\Omega_{bbb}$ excitation energies. For each confinement scenario, interaction parameters are independently optimized against the combined spectra. We find that the Y-shaped potential provides a consistent description of both systems, whereas the $\Delta$-shaped potential exhibits significant discrepancies. Our results provide spectroscopic evidence favoring a collective three-body confining interaction and establish triply-heavy baryon spectroscopy as a complementary probe of the nonperturbative geometry of color confinement.
\end{abstract}
\date{\today}

\maketitle

\emph{Introduction.--}
Understanding how quarks are confined into color-singlet hadrons remains one of the central challenges of quantum chromodynamics (QCD). While the interaction between a heavy quark and antiquark is well described by a Coulomb-like potential at short distances and a linearly rising potential at large separations, the confining interaction among three quarks exhibits a much richer structure. A fundamental question is whether baryonic confinement can be described by a superposition of pairwise interactions or requires a genuine three-body potential. This distinction reflects the non-Abelian nature of QCD and the global or collective organization of color fields in baryons~\cite{Brambilla:2009cd, Brambilla:2005yk}.

\begin{figure}[!htb]
$$\includegraphics[width=0.4\textwidth]{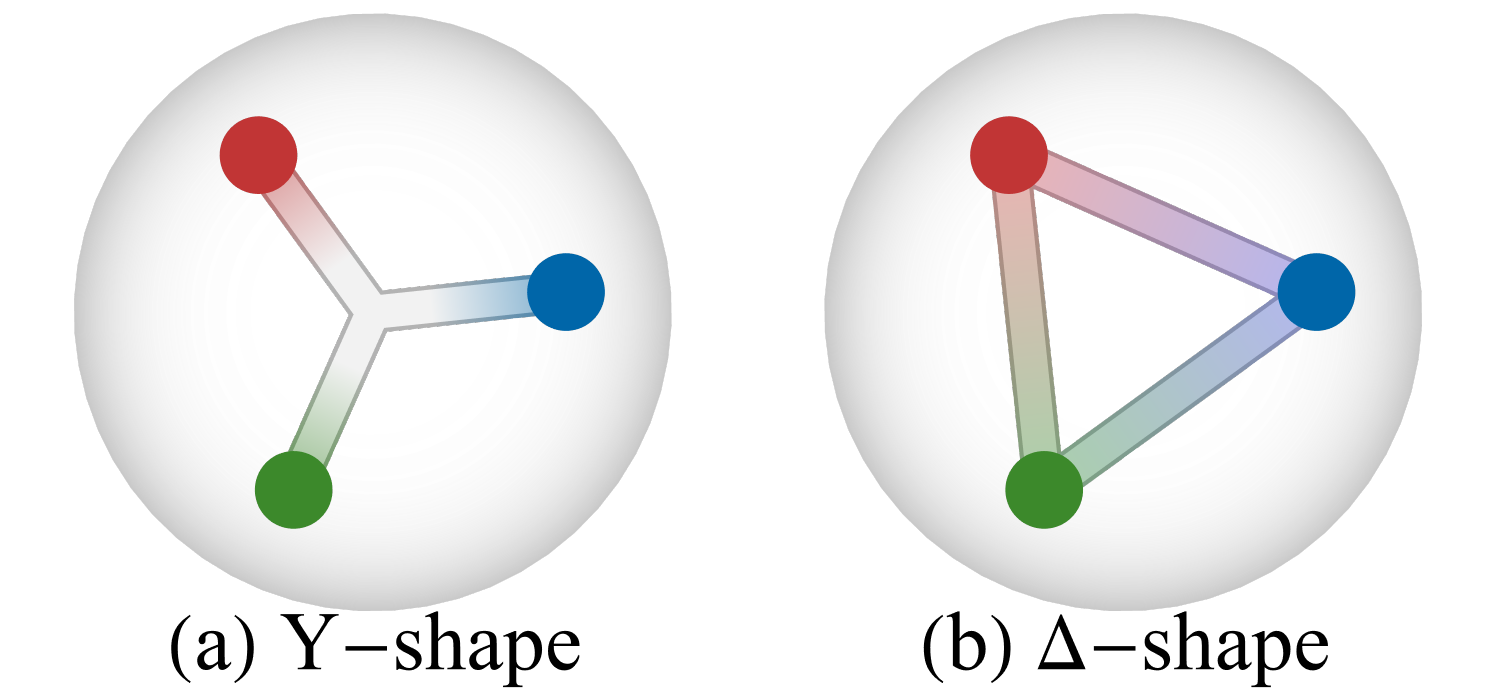}$$
\caption{Illustrations of (a) Y-shaped and (b) $\Delta$-shaped color flux tubes in heavy flavor baryons.}
\label{fig1}
\end{figure}
Two characteristic configurations---respectively referred to as Y and $\Delta$ shapes---have been proposed for the long-range confining interaction among three quarks. In the Y-shaped configuration, three color flux tubes meet at a common junction, and the confining energy is proportional to the minimum total flux-tube length. This collective interaction cannot, in general, be reduced to a sum of pairwise potentials. In contrast, the $\Delta$-shaped configuration approximates confinement by a sum of pairwise interactions, with the confining energy proportional to the perimeter of the triangle~\cite{Silvestre-Brac:1996myf, Martynenko:2007je, Vijande:2015faa}. See Fig.~\ref{fig1} for illustration. Lattice QCD calculations with static external source have provided substantial supports for Y-shaped confinement through the three-quark potential extracted from baryonic Wilson loops~\cite{Takahashi:2000te, Takahashi:2002bw}. Nevertheless, establishing the Y-shaped structure in the static limit does not directly demonstrate how its distinctive geometry manifests itself in the observable spectrum of realistic baryons. In particular, the Y- and $\Delta$-shaped potentials can produce similar binding energies over certain ranges of quark separations, making their spectroscopic discrimination a nontrivial question. Can baryonic excitation spectra resolve these competing confinement geometries and provide experimentally testable signatures of genuine three-body confinement?

Triply-bottom baryons, named $\Omega_{bbb}$, offer a particularly clean theoretical laboratory for addressing this question. The large bottom-quark mass suppresses relativistic corrections, permits a nonrelativistic treatment of the bound-state dynamics, and enables a clear separation between the heavy valence quarks and the light sea quarks. The precisely measured bottomonium spectrum provides stringent constraints on the heavy-quark interaction, whereas lattice QCD has determined the excitation spectrum of $\Omega_{bbb}$, including radial, orbital, and spin-dependent excitations~\cite{Meinel:2012qz}. These complementary data enable a quantitative test of whether the same underlying interaction can simultaneously describe $b\bar{b}$ and $bbb$ systems.
More importantly, identifying excitation patterns that are sensitive to the confinement geometry could establish heavy-baryon spectroscopy as an independent probe of the nonperturbative structure of QCD, providing predictions for future experimental investigations.

In this Letter, we demonstrate that the excitation spectrum of $\Omega_{bbb}$ provides a sensitive spectroscopic probe of baryonic confinement geometry. By solving the two- and three-body Schr\"odinger equations and independently optimizing the interaction parameters against experimentally measured bottomonium masses and lattice QCD calculated $\Omega_{bbb}$ excitation energies, we find that the Y-shaped confinement potential provides a consistent description of both spectra, whereas the pairwise $\Delta$-shaped potential exhibits substantial discrepancies even after a full parameter optimization. Our results establish a connection between the geometry of the confining interaction and the excitation pattern of baryons, providing spectroscopic signatures that may be tested in future experiments.

\vspace{3mm}
\emph{Three-body Schr\"odinger equation.--}
Owing to the large heavy-quark mass, fully heavy-quark systems exhibit a hierarchy of scales, $m_Q\gg m_Qv\gg m_Qv^2$, which enables a systematic nonrelativistic description within effective field theories such as nonrelativistic QCD (NRQCD) and potential NRQCD (pNRQCD)~\cite{Caswell:1985ui,Brambilla:1999xf,Brambilla:2013vx}. In this framework, the interquark potential emerges as a matching coefficient after integrating out the soft scale, providing a theoretical foundation for potential-based descriptions of fully heavy hadrons.
Heavy-quarkonium spectra are well described by a two-body Schr\"odinger equation with a Cornell-type potential~\cite{Zhao:2020jqu}, and the approach can be extended to triply-heavy baryons~\cite{Silvestre-Brac:1996myf,Martynenko:2007je,Vijande:2015faa,Liu:2019vtx,Yang:2019lsg,Shi:2019tji}. For three heavy quarks with mass $m$, the wave function $\Psi(\{{\bm r}\})$---with $\{{\bm r}\}$ being the shorthand of $\{\bm r_1,\bm r_2,\bm r_3\}$---and mass $M$ of their bound state satisfy the three-body Schr\"odinger equation
\begin{equation}
\left[ \sum_{i=1}^3 \left(m+{\hat {\bm p}^2_i\over 2m}\right)+ V(\{{\bm r}\}) \right]\Psi(\{{\bm r}\}) = M\,\Psi(\{{\bm r}\}).
\label{eq.3bodyse}
\end{equation}
The interaction among heavy quarks in a color-singlet baryon consists of a short-range Coulomb component generated by one-gluon exchange between two quarks and a long-range confining interaction associated with color-flux-tube formation. Taking into account the spin-spin interaction which plays an important role for the hyperfine structure of excited states, the interaction potential of the three-quark system reads
\begin{equation}
V(\{{\bm r}\}) = V_\mathrm{conf}(\{{\bm r}\}) + \sum_{i<j}\left[-\frac{\alpha}{r_{ij}} + V_\mathrm{ss}(r_{ij}){\bf s}_i\cdot{\bf s}_j\right],
\label{eq:VD}
\end{equation}
where $r_{ij} = |{\bm r}_i-{\bm r}_j|$ is the relative distance between quarks $i$ and $j$.

At leading order in perturbative QCD, the color factor of the Coulomb interaction between two quarks in a color-singlet $QQQ$ system is one-half of that in a color-singlet $Q\bar Q$ system, $\alpha_{QQ} = \alpha_{Q\bar{Q}}/2$. Thus, the short-distance interaction in triply heavy baryons is expected to be well constrained by the heavy-quarkonium sector.

Spin-dependent interaction is essential for reproducing the hyperfine structure of heavy-hadron spectra.
Lattice QCD studies have extracted the spin-spin interaction in heavy quark-antiquark systems~\cite{Kawanai:2011jt}, motivating a finite-range parametrization of the form $V^{\mathrm{ss}}_{Q\bar Q}(r)=\beta \,e^{-\gamma\, r}$, where $\beta$ is the strength and $\gamma$ the inverse range. Following the same relation between the color factors in the Coulomb interaction, the spin interaction strengths in the $QQQ$ system and $Q\bar Q$ system are related to each other through $\beta_{QQ}=\beta_{Q\bar Q}/2$; meanwhile, the inverse range, which regulates the $\delta$-function-like spin-spin coupling, is taken to be identical in meson and baryon systems, $\gamma_{QQ}=\gamma_{Q\bar Q}$.

Due to its non-perturbative nature, the structure of the confining potential remains an open question. In a $Q\bar{Q}$ system, the confining energy arises from the energy of the color-flux-tube connecting $Q$ and $\bar{Q}$, which leads to a potential of the form $\sigma\,r$. Yet, the configuration of the color-flux-tube in a baryon remains undetermined. It is widely believed that the flux-tube might either form a Y-shape configuration in which the strings start from each quarks and meet at a common junction~\cite{Takahashi:2004rw,Suganuma:2000bi}, as shown in Fig.~\ref{fig1}(a); or a $\Delta$-shape tube that connects the quarks pairwise, illustrated by Fig.~\ref{fig1}(b), and gives a potential~\cite{Silvestre-Brac:1996myf, Martynenko:2007je, Vijande:2015faa}, 
\begin{equation}
V^\mathrm{conf}_\Delta(\{{\bm r}\}) = \sum_{i<j} \sigma \, r_{ij}.
\label{eq:VD}
\end{equation}
Contrastingly, the Y-shaped flux tube gives rise to an intrinsically non-pairwise interaction
\begin{equation}
V_{\mathrm{Y}}^\mathrm{conf}(\{{\bm r}\}) = \sum_i \sigma\, r_{ic},
\label{eq:VY}
\end{equation} 
where $r_{ic} = |{\bm r}_i-{\bm r}_c|$ is the distance from quark $i$ to the flux-tube junction ${\bm r}_c$. The junction location is referred to as the Fermat--Torricelli point~\cite{Takahashi:2002bw}, which minimizing the total string length,
\begin{equation}
L_{\mathrm{Y}}=\sum_i|{\bm r}_i-{\bm r}_c|.
\end{equation}
This minimum-length prescription is implemented explicitly in our numerical calculation. The bottom quark mass $m$ and the potential parameters $\alpha, \beta, \gamma, \sigma$ in the two configurations will be respectively determined by fitting the mass spectrum of fully-bottom meson and baryons. 

To solve the three-body Schr\"odinger equation~\eqref{eq.3bodyse}, we adopt the Jacobi coordinates, see e.g. Ref.~\cite{Zhao:2023qww}, and factorize the three-body function into a plane-wave for the center of mass coordinate, ${\bm X} \equiv ({\bm r}_1+{\bm r}_2+{\bm r}_3)/3$, and motion in the relative coordinates ${\bm x}_1\equiv\sqrt{2}/3[{\bm r}_3-({\bm r}_1+{\bm r}_2)/2]$, ${\bm x}_2\equiv\sqrt{1/6}({\bm r}_2-{\bm r}_1)$, which leads to the wavefunction $\Psi({\bm r}_1,{\bm r}_2,{\bm r}_3)=\Theta({\bm X})\Phi({\bm x}_1,{\bm x}_2)$. Then we introduce the hyperspherical coordinates, the hyperradius $\rho = \sqrt{x_1^2+x_2^2}$ and five hyperangles, to express the six relative coordinates. We further expand the wave function in the hyperspherical harmonic (HH) basis (the spherical harmonic functions correspond to high-dimensional hyperangles), construct the color, spin, and coordinate wave functions subject to the antisymmetry under the exchange of any two identical heavy quarks in a color-singlet baryon. In practical calculations, the infinite HH basis must be truncated. The truncation is performed consistently with the required permutation symmetry by retaining only basis states characterized by the allowed quantum numbers. Numerical convergence is verified by systematically increasing the upper truncation limit. With this expansion, the three-body Schr\"odinger equation~\eqref{eq.3bodyse} yields a set of coupled radial equations, and we solve the equation set with inverse power method~\cite{CRATER1994470, Zhao:2024atb}. Calculation of the potential matrices can be time consuming, but we have adopted the Raynal--Revai transformation between HHs' corresponding to different definitions of Jacobi coordinate and highly improved the efficiency~\cite{Raynal:1970ah}.
The details regarding Jacobi coordinates, HHs, antisymmetric wavefunction construction, and calculation of potential matrices are shown in the Supplemental Materials.

\vspace{3mm}
\emph{Constraining the potential from spectroscopy.--}
The experimental bottomonium spectrum, compiled by the Particle Data Group (PDG)~\cite{10.1093/ptep/ptaa104}, provides stringent constraints on the heavy-quark interaction, including the Coulomb coupling, string tension, bottom-quark mass, and spin-dependent terms. Complementary information on three-body confinement is provided by lattice nonrelativistic QCD (NRQCD) calculations of the triply-bottom baryon $\Omega_{bbb}$ with dynamical light quarks~\cite{Meinel:2012qz}. These calculations determine the excitation spectrum of ten $bbb$ states with $J^P=1/2^+,\,3/2^+,\,5/2^+,\,7/2^+,\,1/2^-$, and $3/2^-$. Combining the two spectra therefore enables a unified determination of the heavy-quark interaction and a direct comparison of competing baryonic confinement geometries.

To consistently compare the calculated spectra with the available data, we employ the experimentally measured bottomonium masses and the lattice-QCD simulated excitation energies of $\Omega_{bbb}$. Since the lattice NRQCD energies contain an unphysical additive offset associated with the heavy-quark rest mass, only energy differences relative to the $J^P=3/2^+$ ground state are extracted. Considering that spin-orbit and tensor interactions are not explicitly included in our Hamiltonian, the corresponding experimental and lattice-QCD multiplets are compared using spin-weighted centroid energies, with weights proportional to $2J+1$.

We determine the five interaction parameters $\boldsymbol{c}=(m,\alpha,\beta,\gamma,\sigma)$ through a simultaneous fit to the bottomonium and $\Omega_{bbb}$ spectra. For each confinement geometry, the two- and three-body Schr\"odinger equations (for bottomonium the Y- and $\Delta$-shaped geometries are reduced to the same flux tube between $Q$ and $\bar Q$) are solved using a common parameter set, and the optimal parameters are obtained by minimizing the $\chi^2$-function
\begin{eqnarray}
\chi^2(\boldsymbol{c})
&=&\chi^2_{bbb}(\boldsymbol{c})+\chi^2_{b\bar b}(\boldsymbol{c})\nonumber\\
&=&\frac{1}{2} \sum_{n\in bbb}
\left[ \frac{\Delta M_n(\boldsymbol{c})-\Delta M_n^{\mathrm{lQCD}}}{\delta\Delta M_n^{\mathrm{lQCD}}} \right]^2\nonumber\\
&&+\frac{1}{2}\sum_{m\in b\bar b}\frac{\left[M_m(\boldsymbol{c})-M_m^{\mathrm{Exp}}\right]^2}{(\delta M_m^{\mathrm{Exp}})^2
+(\delta M_m^{\mathrm{Theo}})^2 },
\label{eq:loss}
\end{eqnarray}
where $\Delta M_n=M_n-M_0$ denotes the $\Omega_{bbb}$ excitation mass relative to its ground state, and $\delta\Delta M_n^{\mathrm{lQCD}}$ is the corresponding lattice uncertainty. The quantities $M_m^{\mathrm{Exp}}$ and $\delta M_m^{\mathrm{Exp}}$ are the experimental bottomonium masses and their uncertainties. To account for the intrinsic limitations of the potential model, we introduce a theoretical uncertainty of $\delta M_m^{\mathrm{Theo}}=20$~MeV for each bottomonium state for the ignorance of spin-orbit and tensor interactions. This prevents the exceptionally small experimental uncertainties from disproportionately dominating the optimization and provides a more balanced constraint from the bottomonium and $\Omega_{bbb}$ spectra. The same uncertainty prescription is applied to both confinement scenarios, ensuring a consistent comparison.
Importantly, all five parameters, $\alpha, \sigma, m, \beta$ and $\gamma$, are determined independently for each confinement scenario through a simultaneous fit to the two- and three-body spectra. This procedure allows both geometries to adjust their short- and long-range interactions on an equal footing, providing a stringent test of their ability to describe heavy-quark spectroscopy within a unified framework. Details of the parameter optimization are given in the Supplemental Materials.

\begin{figure*}[!htb]
	\includegraphics[width=0.56\textwidth]{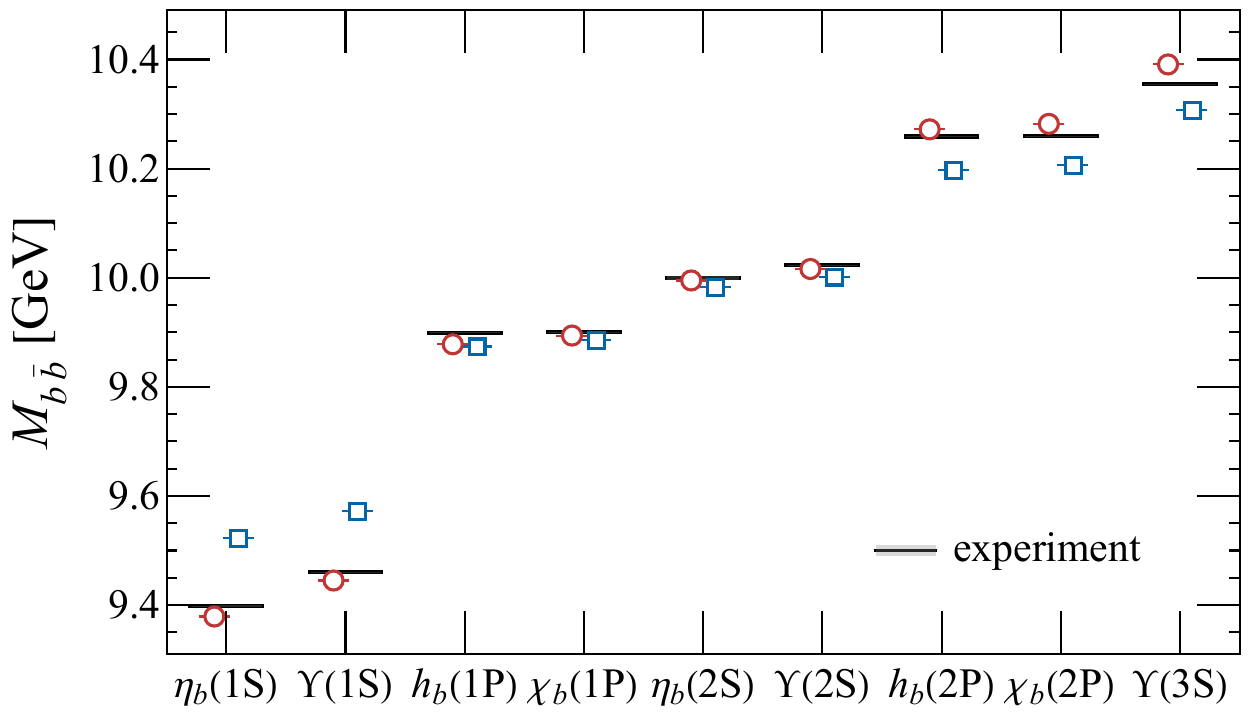}
	\includegraphics[width=0.43\textwidth]{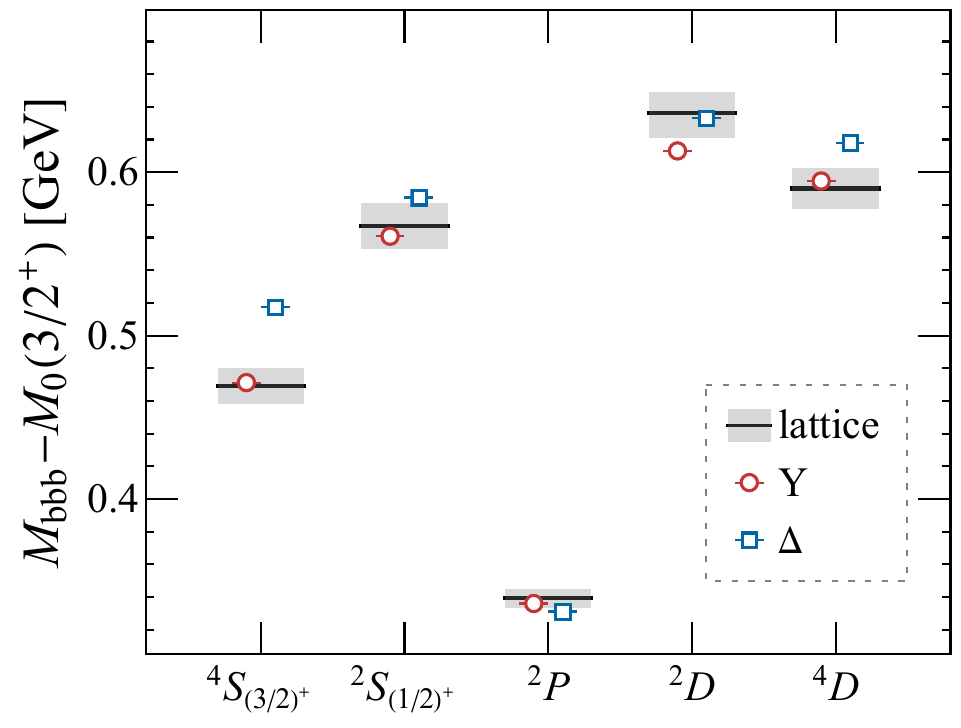}
	\caption{Comparison of the optimized Y-shaped (open red circles) and $\Delta$-shaped (filled purple circles) confinement models with experimentally measured bottomonium masses (left, lines) and lattice QCD calculation of $\Omega_{bbb}$ excitation energies (right, bands). }
	\label{fig2}
\end{figure*}

\vspace{3mm}
\emph{Results.--}
The resulting spectra for the independently optimized Y- and $\Delta$-shaped confinement potentials are compared with experimental bottomonium masses and lattice-QCD $\Omega_{bbb}$ excitation energies in Fig.~\ref{fig2}. The final parameters extracted from the best fit are summarized in Table~\ref{tab1}. 

\begin{table}[!bt]
\renewcommand\arraystretch{2.0}
\caption{Interaction parameters in the Y- and $\Delta$-shaped confinement scenarios, obtained from the best fit to the experimental bottomonium masses and lattice-QCD $\Omega_{bbb}$ excitation energies. The last column are the corresponding minimum values of the $\chi^2$-function.}
\label{tab1}
\setlength{\tabcolsep}{0.8mm}
\begin{tabular}{c|c|c|c|c|c|c}
	\toprule[1pt]\toprule[1pt] 
	& $\alpha$  & $\sigma$(GeV$^2$) & $m$(GeV) & $\beta$(GeV) & $\gamma$(GeV) & $\chi^2_\mathrm{min}$  \\
    \toprule[1pt] 
    $\Delta$ &0.151 & 0.177 & 4.702 & 0.318  & 1.982 & 61.8  \\  
    \toprule[1pt] 
    Y & 0.226 & 0.220 & 4.871 & 0.316 & 2.000 & 5.4\\
	\bottomrule[1pt]\bottomrule[1pt]
\end{tabular}
\end{table}

A striking difference emerges between the two confinement scenarios. The Y-shaped potential can simultaneously reproduce the experimental bottomonium spectrum and the lattice QCD results of $\Omega_{bbb}$ excitation energies with good accuracy, as shown in Fig.~\ref{fig2}. In particular, the predicted baryonic excitation energies closely follow the lattice results across the different orbital and radial excitations. The agreement is achieved using a unified set of interaction parameters, without introducing additional ones specific to the baryonic sector. This demonstrates that a confinement potential governed by the minimal length of a global Y-shaped flux-tube configuration is compatible with both heavy quark-antiquark and three-quark spectroscopy.

In contrast, the $\Delta$-shaped potential exhibits appreciable discrepancies after independently optimizing all five parameters. As shown in Fig.~\ref{fig2}~(right), the calculated $4S_{(3/2)^+}$ excitation energy lies significantly above the lattice-QCD result, while clear deviations also appear for $2S_{(1/2)^+}$ and $^4D$ excitations. Although the $\Delta$ configuration describes well the state $^2D$, it fails to reproduce the complete excitation pattern with reasonable accuracy. Moreover, Fig.~\ref{fig2}~(left) reveals a deterioration in the description of the bottomonium spectrum, particularly for the low-lying $S$-wave states and higher radial excitations. These discrepancies indicate that the pairwise confinement ansatz cannot simultaneously accommodate the two- and three-body spectral constraints, as successfully as the Y-shaped potential.

Quantitative comparison of the fit quality also shows substantial difference: the Y-shaped potential yields a minimum loss of $\chi^2_{\min}=5.4$, compared to $\mathcal \chi^2_{\min}=61.8$ for the $\Delta$-shaped potential. Since both scenarios contain the same five adjustable parameters and are optimized against identical spectral constraints, this marked difference provides quantitative evidence favoring the Y-shaped confinement geometry within the present potential framework.

The simultaneous analysis of bottomonium and $\Omega_{bbb}$ spectra is essential for this discrimination. Bottomonium provides stringent constraints on the heavy-quark mass and the short-range interactions, while the triply-bottom baryon introduces sensitivity to the spatial structure of three-quark confinement. The combined fit therefore restricts the freedom to compensate for an inadequate baryonic confinement geometry through changes in the remaining interaction parameters. The superior agreement obtained with the Y-shaped potential provides spectroscopic evidence favoring a collective three-body confining interaction over a purely pairwise linear ansatz.

\begin{figure}[!h]
    \centering
    \includegraphics[width=0.4\textwidth]{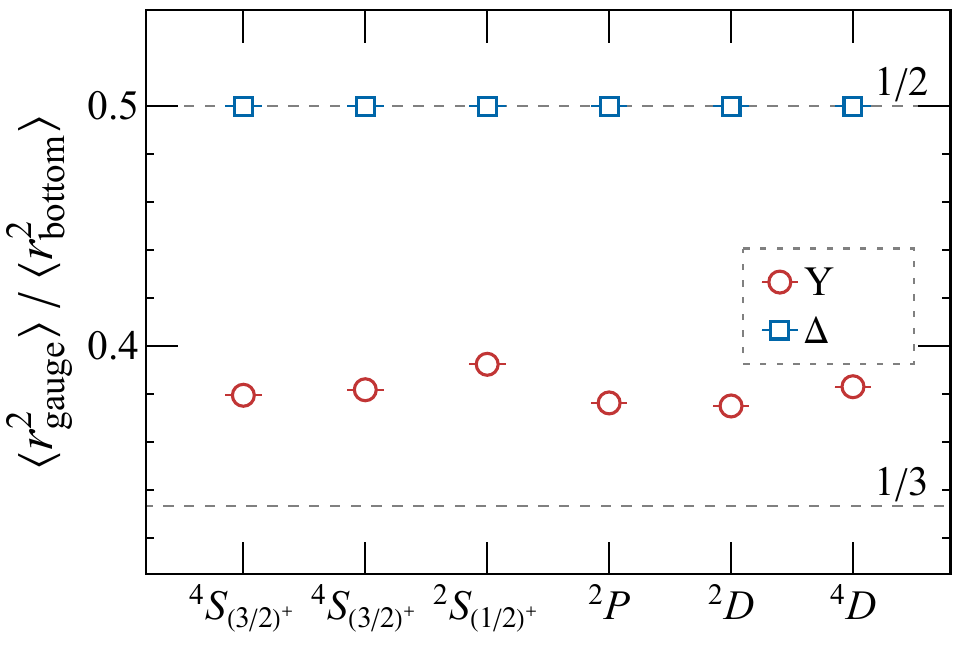}
    \caption{Ratios between the mean-squared radius of the gauge field energy and that of the bottom quarks.
    \label{fig3}}
\end{figure}

While the energy spectrum is dominated by the overall nature of the baryons, a decisive distinction between the Y and $\Delta$ configurations can be made by observables that directly reflect the inner structure.
A good candidate is the ratio between the mean-squared radius of the gauge field energy density distribution and that of the bottom quark. Both of them can be computed in lattice QCD simulations respectively as $\langle r_{\mathrm{gauge}}^2 \rangle \equiv \int r^2 \rho_{\varepsilon,\mathrm{gauge}}(\boldsymbol{r}) \mathrm{d}^3\boldsymbol{r}$ and $\langle r_{\mathrm{bottom}}^2 \rangle \equiv \int r^2 \rho_{n,\mathrm{bottom}}(\boldsymbol{r}) \mathrm{d}^3\boldsymbol{r}$. Neglecting its width, an equilateral-triangular Y-shaped flux tube gives that $\langle r_{\mathrm{gauge}}^2 \rangle / \langle r_{\mathrm{bottom}}^2 \rangle = 1/3$, whereas such a ratio becomes $1/2$ for a $\Delta$ configuration. In Fig.~\ref{fig3} we present these ratios for different $\Omega_{bbb}$ states, computed from the wavefunctions. Good consistency with the equilateral-triangle-based estimations is observed for $\Delta$ configuration, whereas the ratios assuming Y configuration exhibits $\sim 15\%$ deviations---note that non-equilateral-triangle configurations always bring higher the Y-to-bottom ratio, but not the $\Delta$-to-bottom one; see supplemental material for details. Even so, the Y-versus-$\Delta$ difference remains significant and robust against states, and it calls for a lattice-QCD calculation to draw a crucial conclusion.

\vspace{3mm}
\emph{Summary and Discussions.--} 
In the frame of the potential model, we systematically studied the non-perturbative color structures of fully heavy flavor mesons and baryons. We found that the Y-shaped color flux tube corresponding to a direct three-body interaction can fully describe the experimental mass spectrum of mesons and the lattice calculated mass spectrum of baryons, while the $\Delta$-shaped color flux tube which is a simple superposition of pairwise interactions can only reproduce a part of the mass spectrum. This means that the experiment and lattice calculation support the global color structure of Y-shaped configuration. Note that mesons have only a simple color flux tube, only by considering both mesons and baryons can distinguish the non-perturbative structure of strong interaction. 

As the eigenstate of non-perturbative QCD Hamiltonian, the properties of a hadron are determined by its global or topological structure, rather than being controlled by a simple superposition of the constituents. This has been supported by many experimental results and theoretical calculations. For example, the spin of a proton is not the simple summation of the spin of its constituent quarks~\cite{EuropeanMuon:1987isl}. Recent experiment shows also that the baryon number distribution within a nucleon aligns more with that of a global baryon junction, compared to that of the constituent quarks~\cite{STAR:2024lvy}. Our results here further gives the evidence that non-perturbative color interaction is controlled by a global color junction. 

While global quantities such as the mass spectra can clearly distinguish the Y-shaped and $\Delta$-shaped color flux tubes of baryons, more differential quantities such as distribution functions should be more sensitive to the color geometry of baryons. We look forward to future high-luminosity experiment data analysis and high-precision lattice QCD calculations to further identify the non-perturbative inner structure of hadrons~\cite{LiXiaQing:2026}.

\vspace{3mm}
\textbf{Acknowledgment.--}  The authors thank Dr. Kai Zhou for helpful discussions. JZ is supported by the Deutsche Forschungsgemeinschaft (DFG) through the grant CRC-TR 211 ``Strong-interaction matter under extreme conditions'' (Project number 315477589-TRR 211), SS by NSFC under grant No. 12575143, National Key Research and Development Program of China under Contract No. 2024YFA1610700, and Tsinghua University under grant Nos. 04200500123, 531205006 and 533305009, and PZ by Yantai University under the grant No. 2226001.

\clearpage
\begin{appendix}
\begin{widetext}
{\hfill\textbf{SUPPLEMENTAL MATERIAL}\hfill}

\vspace{3mm}
In this Supplemental Material, we present the details of the calculation, which includes Jacobi coordinates and Hyperspherical harmonics (Sec.~\ref{sec:jacobi}), construction of the symmetric spin-spatial basis (Sec.~\ref{sec:color_spin_wf}), the evaluation of the corresponding potential matrix elements (Sec.~\ref{sec:matrix_element}), parameter optimization (Sec.~\ref{sec:parameter_optimization}), and calculation of the mean-squared radius of the gauge field energy (Sec.~\ref{sec:msradius}).

\section{Schroedinger equation with Jacobi coordinates and Hyperspherical harmonics}\label{sec:jacobi}
To solve the three-body Schr\"odinger equation, we introduce the Jacobi coordinates (Set-1),
\begin{eqnarray}
{\bf R} &=& {1\over M}(m_1{\bf r}_1+m_2{\bf r}_2+m_3{\bf r}_3), \nonumber\\
{\bf x}_1&=&\sqrt{(m_1+m_2)m_3 \over (m_1+m_2+m_3)\mu}\left({\bf r}_3-{m_1{\bf r}_1+m_2{\bf r_2}\over m_1+m_2}\right),\nonumber\\
{\bf x}_2&=&\sqrt{{m_1m_2\over (m_1+m_2)\mu}}\left({\bf r}_2-{\bf r}_1\right),
\label{jacobi}
\end{eqnarray}
where $\mu$ is an arbitrary mass scale introduced for convenience and does not affect the physical results. In the numerical calculations, we take $\mu=M=\sum_i m_i$. 
Besides Set-1, two additional Jacobi coordinate systems, denoted as Set-2 and Set-3, can be obtained through the cyclic permutations $1\to2,2\to3, 3\to1$ and $1\to3,2\to1, 3\to2$, respectively.
The explicit form of the Set-2 of Jacobi coordinates is,
\begin{eqnarray}
{\bf R} &=& {1\over M}(m_1{\bf r}_1+m_2{\bf r}_2+m_3{\bf r}_3), \nonumber\\
{\bf x}_1&=&\sqrt{(m_2+m_3)m_1 \over (m_1+m_2+m_3)\mu}\left({\bf r}_1-{m_2{\bf r}_2+m_3{\bf r_3}\over m_2+m_3}\right),\nonumber\\
{\bf x}_2&=&\sqrt{{m_2m_3\over (m_2+m_3)\mu}}\left({\bf r}_3-{\bf r}_2\right),
\end{eqnarray}
and Set-3 of Jacobi coordinates is,
\begin{eqnarray}
{\bf R} &=& {1\over M}(m_1{\bf r}_1+m_2{\bf r}_2+m_3{\bf r}_3), \nonumber\\
{\bf x}_1&=&\sqrt{(m_3+m_1)m_2 \over (m_1+m_2+m_3)\mu}\left({\bf r}_2-{m_3{\bf r}_3+m_1{\bf r_1}\over m_3+m_1}\right),\nonumber\\
{\bf x}_2&=&\sqrt{{m_3m_1\over (m_3+m_1)\mu}}\left({\bf r}_1-{\bf r}_3\right).
\end{eqnarray}

Since the potential depends only on the relative coordinates ${\bf x}_i$, the three-body motion can be factorized into a center-of-mass motion and a relative motion,  $\Psi({\bf r}_1,{\bf r}_2,{\bf r}_3)=\Theta({\bf X})\Phi({\bf x}_1,{\bf x}_2)$. The bound-state properties are therefore fully determined by the relative motion, which reduces the problem to a six-dimensional Schr\"odinger equation. It is convenient to introduce hyperspherical coordinates, defined by the hyperradius
$\rho\equiv\sqrt{x_1^2+x_2^2}$ and the hyperangles $\Omega\equiv\{\alpha_2, \theta_1,\phi_1, \theta_2,\phi_2 \}$, where $\alpha_2 \equiv \arcsin(x_2/ \rho)\in [0,\pi/2]$, and $\{x_i, \theta_i, \phi_i\}$ denote the spherical coordinates of ${\bf x}_i$. In these variables, the Schr\"odinger equation for the intrinsic wave function $\Phi(\rho,\Omega)$ takes the form 
\begin{eqnarray}
\left[ {1\over 2\mu}\left( -{d^2 \over d\rho^2}-{5\over \rho}{d \over d\rho}  + {\widehat {\bf K}_2^2\over \rho^2}\right) + V(\rho, \Omega) \right ]\Phi = E_r \Phi,
\label{eq.relative}
\end{eqnarray}
where $\widehat {\bf K}_2$ is the grand angular momentum operator in six dimensions and $E_r$ denotes the relative energy.
The potential $V(\rho, \Omega)$ depends not only on the hyperradius but also on the hyperangles. In this case, the Schr\"odinger equation~\eqref{eq.relative} cannot be separated into purely radial and angular parts. Instead, the wave function is expanded in the hyperspherical harmonic (HH) basis ${\mathcal Y}_\kappa(\Omega)$, which which are eigenfunctions of $\widehat {\bf K}_2^2$,
\begin{equation}
\widehat {\bf K}_2^2{\mathcal Y}_\kappa(\Omega)=K(K+4){\mathcal Y}_\kappa(\Omega),
\end{equation}
where $\kappa$ denotes the set of quantum numbers $\kappa\equiv (n, l_1, l_2, L, K, M)$. 
The grand angular momentum quantum number is $K=2n+l_1+l_2$ with total orbital angular momentum $\hat L=\hat l_1+\hat l_2$ and projection $M=m_1+m_2$. 
For the three-body system, the HH functions can be written explicitly as,
\begin{eqnarray}
{\mathcal Y}_{\kappa}(\Omega)= \sum_{m_1,m_2} \langle l_1m_1l_2m_2|LM\rangle \rangle  Y_{l_1,m_1}(\theta_1,\phi_1)Y_{l_2,m_2}(\theta_2,\phi_2) \left[ {\mathcal N}_2 (\sin \alpha_2)^{l_2} (\cos \alpha_2)^{l_1} P_{n}^{l_2+{1\over 2}, l_1+
{1\over 2}}(\cos 2 \alpha_2) \right]
\label{eq.hh}
\end{eqnarray}
where $P_n^{a,b}$ denotes the Jacobi polynomial. ${\mathcal N}_2$ is the normalization constant,
\begin{eqnarray}
&&{\mathcal N}_2=\sqrt{ (2K+4)n! \Gamma(n+l_{1}+l_2+2) \over \Gamma(n+l_2+{3\over 2}) \Gamma(n+l_{1}+{3\over 2})}. 
\end{eqnarray}

\section{Color and spin wavefunctions}\label{sec:color_spin_wf}
We next construct the color and spin wave functions subject to the symmetry constraints imposed by the Pauli principle. Since the three heavy quarks are identical fermions, the total wave function must be antisymmetric under the exchange of any two particles. For a triply heavy baryon, the flavor wave function is trivial, and the total wave function can therefore be factorized as, $\Psi=\psi_\mathrm{space}\,\phi_\mathrm{c}\,\chi_\mathrm{s}$, where $\psi_{\rm space}$, $\phi_{\rm c}$, and $\chi_{\rm s}$ denote the spatial, color, and spin components, respectively. To form a color-singlet baryon, the color wave function must belong to the totally antisymmetric representation, $\phi_\mathrm{c}=(QQ)_{\bar 3_c}Q_{3_c}$, which is antisymmetric under quark exchange. Consequently, the combined spin-space wave function must be totally symmetric.
The spin basis follows from group theory, $2\otimes 2 \otimes 2= (3\oplus 1)\otimes 2= 4 \oplus 2\oplus 2$, leading to the orthonormal spin states $\{\chi^s, \chi^a, \chi^{ms}, \chi^{ma}\}$,
\begin{eqnarray}
\chi_{+\frac{3}{2}}^s&=&
    \left|\frac{3}{2},+\frac{3}{2}\right>_{\mathrm{S}} =\left| \uparrow \uparrow \uparrow \right>  \nonumber\\
\chi_{+\frac{1}{2}}^s&=&
    \left|\frac{3}{2},+\frac{1}{2}\right>_{\mathrm{S}} = 3^{-1/2} \Big[ \left| \uparrow \uparrow \downarrow \right> + \left| \uparrow \downarrow \uparrow \right> + \left| \downarrow \uparrow \uparrow \right> \Big] \nonumber\\
\chi_{-\frac{1}{2}}^s&=&
    \left|\frac{3}{2},-\frac{1}{2}\right>_{\mathrm{S}} = 3^{-1/2} \Big[ \left|\downarrow \downarrow \uparrow  \right> + \left| \downarrow \uparrow \downarrow \right> + \left| \uparrow \downarrow \downarrow \right> \Big] \nonumber\\
\chi_{-\frac{3}{2}}^s&=&
    \left|\frac{3}{2},-\frac{3}{2}\right>_{\mathrm{S}} =\left| \downarrow \downarrow \downarrow \right> \nonumber\\
\chi_{+\frac{1}{2}}^{ms}&=&
    \left|\frac{1}{2},+\frac{1}{2}\right>_{\mathrm{MS}} = 6^{-1/2} \Big[ \left| \uparrow \downarrow \uparrow \right> +\left| \downarrow \uparrow \uparrow \right>-2\left| \uparrow \uparrow \downarrow \right> \Big] \nonumber\\
\chi_{-\frac{1}{2}}^{ms}&=&
    \left|\frac{1}{2},-\frac{1}{2}\right>_{\mathrm{MS}} = 6^{-1/2} \Big[ 2\left| \downarrow \downarrow \uparrow \right> - \left| \downarrow \uparrow \downarrow\right> - \left|  \uparrow \downarrow\downarrow \right> \Big] \nonumber\\
\chi_{+\frac{1}{2}}^{ma}&=&
    \left|\frac{1}{2},+\frac{1}{2}\right>_{\mathrm{MA}} = 2^{-1/2} \Big[ \left| \uparrow \downarrow \uparrow \right> - \left| \downarrow \uparrow \uparrow \right> \Big] \nonumber\\
\chi_{-\frac{1}{2}}^{ma}&=&
    \left|\frac{1}{2},-\frac{1}{2}\right>_{\mathrm{MA}} = 2^{-1/2} \Big[ \left|  \uparrow \downarrow\downarrow \right>-  \left| \downarrow \uparrow \downarrow \right>  \Big],
\end{eqnarray}
where $s$, $a$, $ms$, and $ma$ denote the fully symmetric, fully antisymmetric, mixed-symmetric, and mixed-antisymmetric representations, respectively. 

\section{Calculation of potential matrices}\label{sec:matrix_element}
The hyperspherical-harmonic (HH) functions ${\mathcal Y}_{\kappa}$ constitute a complete orthonormal basis for the three-body problem. However, they do not, in general, possess definite symmetry properties under particle exchange. Since the three heavy quarks are identical fermions, the Pauli principle requires the total wave function to be antisymmetric under any permutation. It is therefore necessary to construct symmetry-adapted HH basis states with well-defined transformation properties under the permutation group $S_3$.

We begin by constructing the conventional HH basis and combining it with the spin basis introduced above. States with definite permutation symmetry are then obtained by applying the symmetrization operator
\begin{eqnarray}
\mathcal{S}=\frac{1}{6}\left(
I+P_{12}+P_{13}+P_{23}+P_{123}+P_{132}
\right),
\end{eqnarray}
where the sum runs over all elements of the permutation group $S_3$. Here, $P_{ij}$ denotes the exchange operator of particles $i$ and $j$, while $P_{123}$ and $P_{132}$ represent the cyclic and anticyclic permutations, respectively. The permutation operators act simultaneously on the spatial and spin degrees of freedom according to
$P_{ij}=
P_{ij}^{\rm space}\otimes P_{ij}^{\rm spin}$. 

In practical calculations, the infinite HH basis must be truncated. The truncation is performed consistently with the required permutation symmetry by retaining only basis states characterized by the allowed quantum numbers $n$, $l_i$, $L$, and $K$. The resulting symmetry-adapted space $\otimes$ spin basis provides a systematically improvable representation of the three-body Hilbert space while preserving the exact exchange symmetry of the system. Using this basis, the relative wave function of the triply heavy baryon can be expanded as
\begin{eqnarray}
 \Phi(\rho, \Omega)&=&\sum_\kappa R_{\kappa}(\rho)|\phi^{sym}\rangle,
\end{eqnarray}
where $\phi^{sym}$ is the symmetry-adapted spatial-spin basis.
Substituting this expansion into the relative Schr\"odinger equation~\eqref{eq.relative} yields a set of coupled radial equations,
\begin{equation} 
\left[{1\over 2\mu}\left({d^2 \over d\rho^2}+{5\over \rho}{d \over d\rho} - {K(K+4)\over \rho^2}\right)+E_r \right]R_{\kappa}=\sum_{\kappa'}V_{\kappa \kappa'}R_{\kappa'},
\end{equation}
For $\Delta$-shape, the matrix elements of the pairwise potential between particles $(i,j)$ can be evaluated efficiently using the Raynal--Revai (RR) transformation coefficients~\cite{Raynal:1970ah}, which relate hyperspherical-harmonic basis functions defined in different Jacobi coordinate systems.
\begin{eqnarray}
V_{\kappa\kappa'}^{(ij)} (\rho) &\equiv&\int V^{(ij)} (|\mathbf{r}_j-\mathbf{r}_i|) {\mathcal Y}_\kappa^*(\Omega) {\mathcal Y}_{\kappa'}(\Omega)\mathrm{d}\Omega \nonumber\\
&=& \sum_{kk'} (R^{(ij)}_{\kappa k})^{*} R^{(ij)}_{\kappa' k'} 
	\int V^{(ij)} (|\mathbf{r_j}-\mathbf{r_i}|){\mathcal Y}_k^*(\Omega^{(ij)}){\mathcal Y}_{k'}(\Omega^{(ij)})\mathrm{d}\Omega^{(ij)} \nonumber\\
&=&\sum_{kk'} (R^{(ij)}_{\kappa k})^{*} R^{(ij)}_{\kappa' k'} V_{kk'}^{(12)}(\rho).
\end{eqnarray}
In the Jacobi coordinate Set-1 where particles 1 and 2 are paired, the interaction depends only on the hyperangle $\alpha_2$, leading to a significant simplification,
\begin{eqnarray}
V_{\kappa\kappa'}^{(12)} &\equiv& \int V^{(12)}(|\mathbf{r}_2-\mathbf{r}_1|) 
	{\mathcal Y}_\kappa^*(\Omega){\mathcal Y}_{\kappa'}(\Omega)\mathrm{d}\Omega \\
&=&\int V^{(12)}(\sqrt{2\mu/m}\, \rho \sin \alpha_2)
	{\mathcal Y}_\kappa^*(\Omega){\mathcal Y}_{\kappa'}(\Omega)\mathrm{d}\Omega, \nonumber
\end{eqnarray}
with the hyperspherical volume element
\begin{equation}
d\Omega= \cos^2 \alpha_2 \sin^2 \alpha_2 \sin \theta_1 \sin \theta_2 d\alpha_2d\theta_1d\phi_1d\theta_2d\phi_2.
\end{equation} 

For the Y-shaped interaction, an analogous procedure is applied for the color--junction terms. The corresponding matrix elements read
\begin{eqnarray}
V_{\kappa\kappa'}^{(ic)} (\rho) &\equiv&\int V^{(ic)} (|\mathbf{r}_{ic}|) {\mathcal Y}_\kappa^*(\Omega) {\mathcal Y}_{\kappa'}(\Omega)\mathrm{d}\Omega \nonumber\\
&=& \sum_{kk'} (R^{(ij)}_{\kappa k})^{*} R^{(ij)}_{\kappa' k'} 
	\int V^{(ic)} (|\mathbf{r}_{ic}|){\mathcal Y}_k^*(\Omega^{(ij)}){\mathcal Y}_{k'}(\Omega^{(ij)})\mathrm{d}\Omega^{(ij)} \nonumber\\
&=&\sum_{kk'} (R^{(ij)}_{\kappa k})^{*} R^{(ij)}_{\kappa' k'} V_{kk'}^{(3c)}(\rho).
\end{eqnarray}
The potential of particle 3 to the Fermat--Torricelli point is highly reduced, 
\begin{eqnarray}
V_{\kappa\kappa'}^{(3c)}(\rho) &\equiv& \int V^{(3c)}(|\mathbf{r}_{3c}|) 
	{\mathcal Y}_\kappa^*(\Omega){\mathcal Y}_{\kappa'}(\Omega)\mathrm{d}\Omega \\
&=&\int V^{(3c)}(\sqrt{2\mu \over 3m}\, \rho \cos \alpha_2)
	{\mathcal Y}_\kappa^*(\Omega){\mathcal Y}_{\kappa'}(\Omega)\mathrm{d}\Omega,\nonumber
\end{eqnarray}
where the potential $V^{(3c)}$ depends only on $\alpha_2$, again allowing a substantial simplification of the angular integrations.

Now, we apply the above mentioned symmetrization operator to generate the symmetric spatial-spin basis for each state. Here, we take $P$-wave states as an example. For $1^2P_{{1\over 2}^-}$ and $1^2P_{{3\over 2}^-}$ state, the HH basis should satisfy the constraints: $l_1+l_2=\rm odd$ for $P=-1$.
$S={1\over 2}$, $L=1$, $J={1\over 2}, {3\over 2}$.
In the numerical calculation, we need to make a truncation according to the symmetry properties of the system. Therefore, for $1^2P_{{1\over 2}^-}$ or $1^2P_{{3\over 2}^-}$ state, we take the basis for $K\le 4$,
\begin{eqnarray}
&&{\mathcal Y}_{(0,1,0,1,1,0)}={\sqrt{6}\over \pi^{3/2}}\cos\alpha_2\cos \theta_1, \nonumber\\
&&{\mathcal Y}_{(0,0,1,1,1,0)}={\sqrt{6}\over \pi^{3/2}}\sin\alpha_2\cos \theta_2, \nonumber\\
&&{\mathcal Y}_{(1,1,0,1,3,0)}={\sqrt{2}\over \pi^{3/2}}(\cos\alpha_2+2\cos3\alpha_2)\cos\theta_1,  \nonumber\\
&&{\mathcal Y}_{(1,0,1,1,3,0)}={\sqrt{2}\over \pi^{3/2}}(1+4\cos2\alpha_2)\sin\alpha_2\cos\theta_2, \nonumber\\
&&{\mathcal Y}_{(0,1,2,1,3,0)}={-2\over \pi^{3/2}}\sin^2\alpha_2\cos\alpha_2(\cos\theta_1+3\cos2\theta_2\cos \theta_1+3\cos(\phi_1-\phi_2)\sin\theta_1\sin2\theta_2), \nonumber\\
&&{\mathcal Y}_{(0,2,1,1,3,0)}={-2\over \pi^{3/2}}\cos^2\alpha_2\sin\alpha_2(\cos\theta_2+3\cos2\theta_1\cos \theta_2+3\cos(\phi_1-\phi_2)\sin2\theta_1\sin\theta_2).
\end{eqnarray}
where the subscript of $\mathcal{Y}$ denotes the set of quantum numbers $\kappa\equiv (n, l_1, l_2, L, K, M)$ as mentioned before.

We first compute the $P_{ij}$. $K$ and $L$ are preserved under any permutation $P_{ij}$.
Since the permutation of particle 1 and 2 affects only the last Jacobi coordinate for set-1,
\begin{eqnarray}
\hat P_{12} ({\bf x}_1, {\bf x}_2)= ({\bf x}_1, -{\bf x}_2),
\end{eqnarray}
where $\hat P_{12}$ is the operator associate with the permutation of particles 1 and 2, and since,
\begin{eqnarray}
Y_{l_2,m_2}(-{\bf \hat x}_2)=(-1)^{l_2}Y_{l_2,m_2}(-{\bf \hat x}_2),
\end{eqnarray}
where ${\bf \hat x}_2$ designates the angular part of the spherical coordinates of the ${\bf x}_2$. Therefore, we obtain,
\begin{eqnarray}
\hat P_{12} {\mathcal Y}_\kappa(\Omega)=(-1)^{l_2}{\mathcal Y}_\kappa(\Omega).
\end{eqnarray} 
That means HH are symmetric with respect to the exchange of particle 1 and 2 if $l_2$ is even and anti-symmetric if $l_2$ is odd. 
$P_{23}$ is the diagonal matrix of set-2, and $P_{13}$ of set-3.
So, we have,
\begin{align}
P_{12}^{(1)} = \left(
\begin{array}{cccccc}
1 & 0 & 0 & 0 & 0 & 0 \\
0 & -1 & 0 & 0 & 0 & 0 \\
0 & 0 & 1 & 0 & 0 & 0 \\
0 & 0 & 0 & -1 & 0 & 0 \\
0 & 0 & 0 & 0 & 1 & 0 \\
0 & 0 & 0 & 0 & 0 & -1 
\end{array}
\right).
\end{align}

\begin{align}
\left(\begin{array}{c}
{\bf x}_1^{(12)} \\
{\bf x}_2^{(12)} \\
\end{array}\right)
= \left(\begin{array}{cc}
a_{11}& a_{12} \\
a_{21}& a_{22} \\
\end{array}\right)
\left(\begin{array}{c}
{\bf x}_1^{(ij)} \\
{\bf x}_2^{(ij)} \\
\end{array}\right)
\end{align}
The definition of the Raynal--Revai (RR) coefficient~\cite{Raynal:1970ah} is 
\begin{eqnarray}
{\mathcal Y}_j^{(1)}(\Omega_1)=\sum_{k} R_{jk} {\mathcal Y}_k^{(2)}(\Omega_2),
\end{eqnarray}
set-2 is the original Jacobi basis, set-1 is the new one.
\begin{eqnarray}
P_{23,ij}^{(1)}&\equiv& \int d\Omega {\mathcal Y}_i^{(1)*}(\Omega_1)P_{23}{\mathcal Y}_j^{(1)}(\Omega_1) \nonumber\\
&=&\int d\Omega \left(\sum_n R_{im}{\mathcal Y}_m^{(2)*}(\Omega_2)\right)^*P_{23}\left(\sum_n R_{jn}{\mathcal Y}_n^{(2)}(\Omega_2)\right)\nonumber\\
&=&\sum_{m,n}R_{im}P_{23,mn}^{(2)}R_{jn} ,
\end{eqnarray}
and in a compact form,
\begin{eqnarray}
P_{23}^{(1)}=RP_{23}^{(2)}R^T.
\end{eqnarray}
This relation is much helpful for calculating the other $P_{ij}$. For $P_{23}$, we can first calculate the $P_{23}^{(2)}$ in the set-2, which is, 
\begin{align}
P_{23}^{(2)} = \left(
\begin{array}{cccccc}
1 & 0 & 0 & 0 & 0 & 0 \\
0 & -1 & 0 & 0 & 0 & 0 \\
0 & 0 & 1 & 0 & 0 & 0 \\
0 & 0 & 0 & -1 & 0 & 0 \\
0 & 0 & 0 & 0 & 1 & 0 \\
0 & 0 & 0 & 0 & 0 & -1 
\end{array}
\right).
\end{align}
Then we need to calculate the RR matrix from set-2 to set-1. Finally, we get,
\begin{align}
P_{23}^{(1)} = \left(
\begin{array}{cccccc}
-{1\over 2} & -{\sqrt{3}\over 2} & 0 & 0 & 0 & 0 \\
-{\sqrt{3}\over 2} & {1\over 2} & 0 & 0 & 0 & 0 \\
0 & 0 & {1\over 2} & {1\over 2\sqrt{3}} & -{1\over \sqrt{2}} & -{1\over \sqrt{6}} \\
0 & 0 & {1\over 2\sqrt{3}} & -{1\over 2} & {1\over \sqrt{6}} & -{1\over \sqrt{2}} \\
0 & 0 & -{1\over \sqrt{2}} & {1\over \sqrt{6}} & 0 & -{1\over \sqrt{3}} \\
0 & 0 & -{1\over \sqrt{6}} & -{1\over \sqrt{2}} & -{1\over \sqrt{3}} & 0 
\end{array}
\right).
\end{align}
Because $K$ and $L$ are conserved under permutation and rotation, the matrix can be dealt in different $K$ block. 
For $P_{13}$, we can first calculate the $P_{13}^{(3)}$ in the set-3, which is, 
\begin{align}
P_{13}^{(3)} = \left(
\begin{array}{cccccc}
1 & 0 & 0 & 0 & 0 & 0 \\
0 & -1 & 0 & 0 & 0 & 0 \\
0 & 0 & 1 & 0 & 0 & 0 \\
0 & 0 & 0 & -1 & 0 & 0 \\
0 & 0 & 0 & 0 & 1 & 0 \\
0 & 0 & 0 & 0 & 0 & -1 
\end{array}
\right).
\end{align}
With the RR matrix from set-3 to set-1, we can get
\begin{align}
P_{13}^{(1)} = \left(
\begin{array}{cccccc}
-{1\over 2} & {\sqrt{3}\over 2} & 0 & 0 & 0 & 0 \\
{\sqrt{3}\over 2} & {1\over 2} & 0 & 0 & 0 & 0 \\
0 & 0 & {1\over 2} & -{1\over 2\sqrt{3}} & -{1\over \sqrt{2}} & {1\over \sqrt{6}} \\
0 & 0 & -{1\over 2\sqrt{3}} & -{1\over 2} & -{1\over \sqrt{6}} & -{1\over \sqrt{2}} \\
0 & 0 & -{1\over \sqrt{2}} & -{1\over \sqrt{6}} & 0 & {1\over \sqrt{3}} \\
0 & 0 & {1\over \sqrt{6}} & -{1\over \sqrt{2}} & {1\over \sqrt{3}} & 0 
\end{array}
\right).
\end{align}
With the relation $P_{123}=P_{12}P_{23}$, and $P_{132}=P_{23}P_{12}$, we have,
\begin{align}
P_{123}^{(1)} = \left(
\begin{array}{cccccc}
-{1\over 2} & -{\sqrt{3}\over 2} & 0 & 0 & 0 & 0 \\
{\sqrt{3}\over 2} & -{1\over 2} & 0 & 0 & 0 & 0 \\
0 & 0 & {1\over 2} & {1\over 2\sqrt{3}} & -{1\over \sqrt{2}} & -{1\over \sqrt{6}} \\
0 & 0 & -{1\over 2\sqrt{3}} & {1\over 2} & -{1\over \sqrt{6}} & {1\over \sqrt{2}} \\
0 & 0 & -{1\over \sqrt{2}} & {1\over \sqrt{6}} & 0 & -{1\over \sqrt{3}} \\
0 & 0 & {1\over \sqrt{6}} & {1\over \sqrt{2}} & {1\over \sqrt{3}} & 0 
\end{array}
\right),
\end{align}
\begin{align}
P_{132}^{(1)} = \left(
\begin{array}{cccccc}
-{1\over 2} & {\sqrt{3}\over 2} & 0 & 0 & 0 & 0 \\
-{\sqrt{3}\over 2} & -{1\over 2} & 0 & 0 & 0 & 0 \\
0 & 0 & {1\over 2} & -{1\over 2\sqrt{3}} & -{1\over \sqrt{2}} & {1\over \sqrt{6}} \\
0 & 0 & {1\over 2\sqrt{3}} & {1\over 2} & {1\over \sqrt{6}} & {1\over \sqrt{2}} \\
0 & 0 & -{1\over \sqrt{2}} & -{1\over \sqrt{6}} & 0 & {1\over \sqrt{3}} \\
0 & 0 & -{1\over \sqrt{6}} & {1\over \sqrt{2}} & -{1\over \sqrt{3}} & 0 
\end{array}
\right).
\end{align}

For the spin $S=1/2$, there are two basis $\{\chi_{ms},\chi_{ma}\}$,
\begin{align}
P_{12} = \left(
\begin{array}{cc}
1 & 0  \\
0 & -1 
\end{array}
\right),
P_{13} = \left(
\begin{array}{cc}
-{1\over 2} & {\sqrt{3}\over 2}  \\
{\sqrt{3}\over 2} & {1\over 2} 
\end{array}
\right),
P_{23} = \left(
\begin{array}{cc}
-{1\over 2} & -{\sqrt{3}\over 2}  \\
-{\sqrt{3}\over 2} & {1\over 2} 
\end{array}
\right), \\
P_{123} = \left(
\begin{array}{cc}
-{1\over 2} & -{\sqrt{3}\over 2}  \\
{\sqrt{3}\over 2} & -{1\over 2} 
\end{array}
\right),
P_{132} = \left(
\begin{array}{cc}
-{1\over 2} & {\sqrt{3}\over 2}  \\
-{\sqrt{3}\over 2} & -{1\over 2} 
\end{array}
\right).
\end{align}
The total basis is $P_{ij}=P_{ij}^{\rm space} \otimes P_{ij}^{\rm spin}$. The symmetrizer $\mathcal{S}$ is $12\times 12$ dimensional matrix. For the matrix $\mathcal{S}$, we get eigenvalues $\{1,1,0,0,0,0,0,0,0,0,0,0\}$. There are two non-zero eigenvalues, which means there are exactly two independent orthonormal states that are totally symmetric under all particle permutations. 
The related two eigenstates are:
\begin{eqnarray}
\phi_1^{sym}&=&{1\over \sqrt{2}}\left (\mathcal{Y}_{(0,1,0,1,1,0)}\chi_{ms}+\mathcal{Y}_{(0,0,1,1,1,0)}\chi_{ma}\right),\nonumber\\
\phi_2^{sym}&=&{1\over \sqrt{3}}\Big ({1\over \sqrt{2}}\mathcal{Y}_{(1,1,0,1,3,0)}\chi_{ms}-{1\over \sqrt{2}}\mathcal{Y}_{(1,0,1,1,3,0)}\chi_{ma} +\mathcal{Y}_{(0,1,2,1,3,0)}\chi_{ms}+\mathcal{Y}_{(0,2,1,1,3,0)}\chi_{ma}\Big).
\end{eqnarray}
Now, we can expand the total relative wave function, for example,
 \begin{eqnarray}
 \Phi^{P}(\rho, \Omega)&=&\sum_\kappa R_{\kappa}(\rho)|\phi^{sym}\rangle \nonumber\\
 &=&R_{1}(\rho)\phi_1^{sym}(\Omega,\sigma)+R_{2}(\rho)\phi_2^{sym}(\Omega,\sigma).
\end{eqnarray}
It's largely suppressed to 2 dimension.

For $1^4D_{{3\over 2}^+}$, $1^4D_{{5\over 2}^+}$, and $1^4D_{{7\over 2}^+}$ states.
The positive parity requires $l_1+l_2=\rm even$. $S={3\over 2}$ and $L=2$ gives $J={3\over 2}, {5\over 2}, {7\over 2}$.
So, the spin wavefunction is fully symmetric. 
The space part can be symmetrized by act $\mathcal{S}$ on the HH basis.
Therefore, the symmetrizer $\mathcal{S}$ is a $9\times 9$ dimensional matrix. The eigenvalues are $\{1,1,0,...,0\}$. Two non-zero eigenvalues give the eigenstates,
\begin{eqnarray}
\phi_1^{sym}&=&{1\over \sqrt{2}}\left (\mathcal{Y}_{(0,2,0,2,2,0)}\chi_{s}+\mathcal{Y}_{(0,0,2,2,2,0)}\chi_{s}\right),\nonumber\\
\phi_2^{sym}&=&{1\over \sqrt{12}}\Big (-\sqrt{10}\mathcal{Y}_{(0,2,2,2,4,0)}\chi_{s}-\mathcal{Y}_{(1,2,0,2,4,0)}\chi_{s}+\mathcal{Y}_{(1,0,2,2,4,0)}\chi_{s}\Big).
\end{eqnarray}
Now, we can expand the total relative wave function,
 \begin{eqnarray}
 \Phi^D(\rho, \Omega)&=&R_{1}(\rho)\phi_1^{sym}(\Omega,\sigma)+R_{2}(\rho)\phi_2^{sym}(\Omega,\sigma).
\end{eqnarray}

$1^2D_{{3\over 2}^+}$ and $1^2D_{{5\over 2}^+}$ state. First construct the basis for the $P_{12}$, which means the symmetric state for exchanging particle 1 and 2, which requires the HH basis satisfies the constraints: $l_2+s_{12}=\rm odd$. $l_1+l_2=\rm even$ is need for positive parity. $S={1\over 2}$ and $L=2$ gives $J={3\over 2}, {5\over 2}$. We have the basis (marking $(n,l_1,l_2,m)$),
Therefore, the symmetrizer $\mathcal{S}$ is a $18\times 18$ dimensional matrix. The eigenvalues are $\{1,1,1,0,...,0\}$. Three non-zero eigenvalues give the eigenstates, $\phi_1^{sym}$, $\phi_2^{sym}$, and $\phi_3^{sym}$. 
These three are linearly independent basis but not orthonormal basis for $\phi_2^{sym}$ and $\phi_3^{sym}$. In principle, linearly independent basis are ok but orthonormal basis are better. The Gram--Schmidt is used to form a orthonormal basis by replace the $\phi_3^{sym}$ with $\tilde \phi_3^{sym}/||\tilde \phi_3^{sym}||$ where $\tilde \phi_3^{sym}=\phi_3^{sym}-\langle \phi_2^{sym}|\phi_3^{sym} \rangle \phi_2^{sym}$,
\begin{eqnarray}
\phi_1^{sym}&=&{1\over 2}\Big (\sqrt{2}\mathcal{Y}_{(0,1,1,2,2,0)}\chi_{ma}-\mathcal{Y}_{(0,2,0,2,2,0)}\chi_{ms}+\mathcal{Y}_{(0,0,2,2,2,0)}\chi_{ms}\Big),\nonumber\\
\phi_2^{sym}&=&\sqrt{5\over 11}\Big ({1\over \sqrt{10}}\mathcal{Y}_{(0,2,2,2,4,0)}\chi_{ms}-{4\over 5}\mathcal{Y}_{(0,3,1,2,4,0)}\chi_{ma} -{1\over 5}\mathcal{Y}_{(0,1,3,2,4,0)}\chi_{ma}-{\sqrt{21/2}\over 5}\mathcal{Y}_{(1,1,1,2,4,0)}\chi_{ma} \nonumber\\
&+&\mathcal{Y}_{(1,0,2,2,4,0)}\chi_{ms}\Big),\nonumber\\
\phi_3^{sym}&=&{1\over 2\sqrt{66}}\Big (-\sqrt{10}\mathcal{Y}_{(0,2,2,2,4,0)}\chi_{ms}-3\mathcal{Y}_{(0,3,1,2,4,0)}\chi_{ma}-9\mathcal{Y}_{(0,1,3,2,4,0)}\chi_{ma}+\sqrt{42}\mathcal{Y}_{(1,1,1,2,4,0)}\chi_{ma}\nonumber\\
&+&11\mathcal{Y}_{(1,2,0,2,4,0)}\chi_{ms}+\mathcal{Y}_{(1,0,2,2,4,0)}\chi_{ms}\Big).
\end{eqnarray}

Now, we can expand the total relative wave function, for example,
 \begin{eqnarray}
 \Phi^D(\rho, \Omega)&=&R_{1}(\rho)\phi_1^{sym}(\Omega,\sigma)+R_{2}(\rho)\phi_2^{sym}(\Omega,\sigma)+R_{3}(\rho)\phi_3^{sym}(\Omega,\sigma).
\end{eqnarray}

For $2^2S_{{1\over 2}^+}$ state, the HH basis should satisfy the constraints: $l_2+s_{12}=\rm odd$. The parity requires $l_1+l_2=\rm even$. $S={1\over 2}$ and $L=0$ gives $J={1\over 2}$. The possible basis are (marking $(n,l_1,l_2,m)$),
Two non-zero eigenvalues give the eigenstates,
\begin{eqnarray}
\phi_1^{sym}&=&{1\over \sqrt{2}}\left (-\mathcal{Y}_{(1,0,0,0,2,0)}\chi_{ms}+\mathcal{Y}_{(0,1,1,0,2,0)}\chi_{ma}\right),\nonumber\\
\phi_2^{sym}&=&{1\over \sqrt{6}}\Big (-\sqrt{2}\mathcal{Y}_{(2,0,0,0,4,0)}\chi_{ms}-\sqrt{3}\mathcal{Y}_{(1,1,1,0,4,0)}\chi_{ma}+\mathcal{Y}_{(0,2,2,0,4,0)}\chi_{ms}\Big).
\end{eqnarray}
Now, we can expand the total relative wave function via,
 \begin{eqnarray}
 \Phi^S(\rho, \Omega)=R_{1}(\rho)\phi_1^{sym}(\Omega,\sigma)+R_{2}(\rho)\phi_2^{sym}(\Omega,\sigma).
\end{eqnarray}

For $1^4S_{{3\over 2}^+}$ (or  $2^4S_{{3\over 2}^+}$, which is radial excited state) state.
$S={3\over 2}$ and $L=0$ gives $J={3\over 2}$.
So, the spin space is fully symmetric. So, we just need to construct a symmetric spatial part. We can act the symmetrizer $\mathcal{S}$ on the HH basis. Therefore, the symmetrizer $\mathcal{S}$ is a $6\times 6$ dimensional matrix.
The eigenvalues are $\{1,1,0,0,0,0\}$. Two non-zero eigenvalues give the eigenstates,
\begin{eqnarray}
\phi_1^{sym}&=&\mathcal{Y}_{(0,0,0,0,0,0)}\chi_{s},\nonumber\\
\phi_2^{sym}&=&\sqrt{1\over 3}\mathcal{Y}_{(2,0,0,0,4,0)}\chi_{s}+\sqrt{2\over 3}\mathcal{Y}_{(0,2,2,0,4,0)}\chi_{s}.
\end{eqnarray}
The total relative wave function is expanded,
\begin{eqnarray}
 \Phi^{S}(\rho, \Omega)=R_{1}(\rho)\phi_1^{sym}(\Omega,\sigma)+R_{2}(\rho)\phi_2^{sym}(\Omega,\sigma).
\end{eqnarray}

In the numerical calculations, we also examined the convergence of each state by systematically enlarging the basis set. The resulting energies were found to be stable, demonstrating the numerical convergence of the basis expansion.

\section{Parameter optimization}\label{sec:parameter_optimization}
\begin{figure}[!htb]
$$\includegraphics[width=0.8\textwidth]{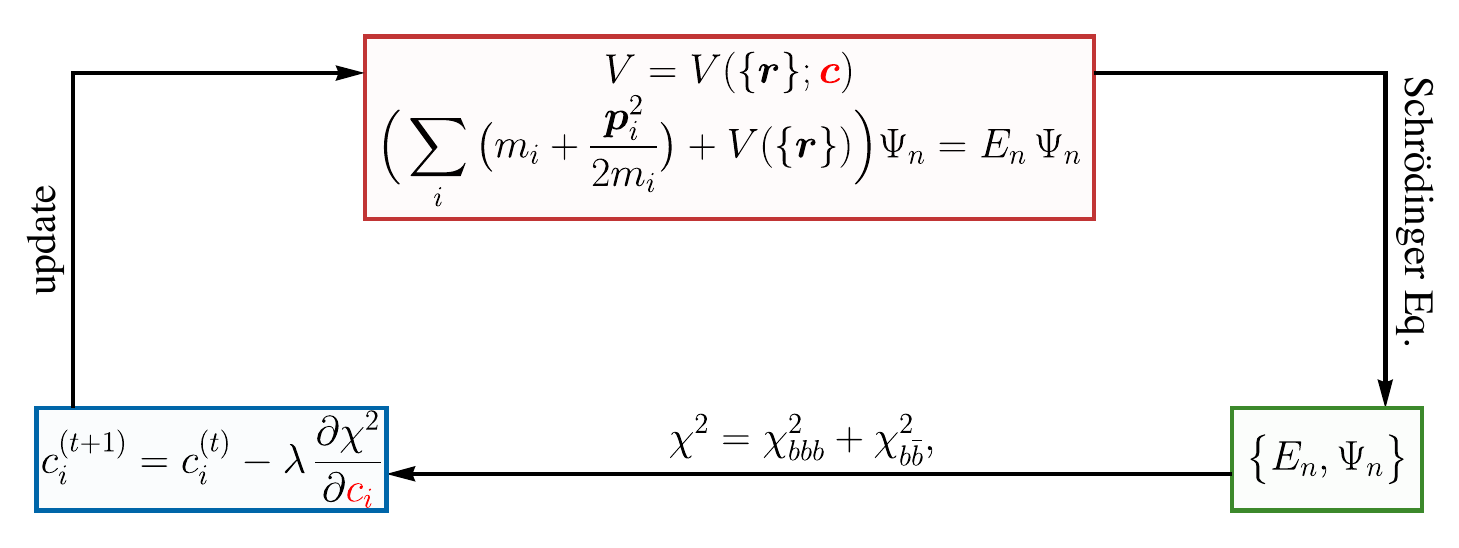}$$
\caption{Flow chart of the potential extraction scheme.}
\label{fig2}
\end{figure}
In this work, the parameters are optimized iteratively using gradient descent, 
\begin{align}
    c_i^{(t+1)} = c_i^{(t)} - \lambda \frac{\partial \chi^2}{\partial c_i},
\end{align}
where $c_i^{(t)}$ is the $i$-th parameter in the $t$-th optimization step, and $\lambda$ is the learning rate, $\chi^2$ is the loss function. See Fig.~\ref{fig2} for illustration.
The derivatives of the energy eigenvalues evaluated through the Feynman--Hellmann theorem,
\begin{equation}
\frac{\partial M_n}{\partial c_i}=
\left\langle\Phi_n\left|
\frac{\partial H}{\partial c_i}
\right|\Phi_n\right\rangle,
\label{eq:FH}
\end{equation}
where $\Phi_n$ is the corresponding eigenstate. For the $\Omega_{bbb}$ excitation masses, the derivatives are obtained by subtracting the ground-state contribution,
\begin{equation}
\frac{\partial\Delta M_n}{\partial c_i} = \frac{\partial M_n}{\partial c_i} - \frac{\partial M_0}{\partial c_i}.
\label{eq:FH_exc}
\end{equation}
The same optimization procedure is applied independently to the Y- and $\Delta$-shaped confinement scenarios, allowing their ability to reproduce the combined two- and three-body spectra to be assessed on an equal footing.

\section{Mean-squared radius of gauge field energy distribution}\label{sec:msradius}
\begin{figure}
    \centering
    \includegraphics[width=0.5\textwidth]{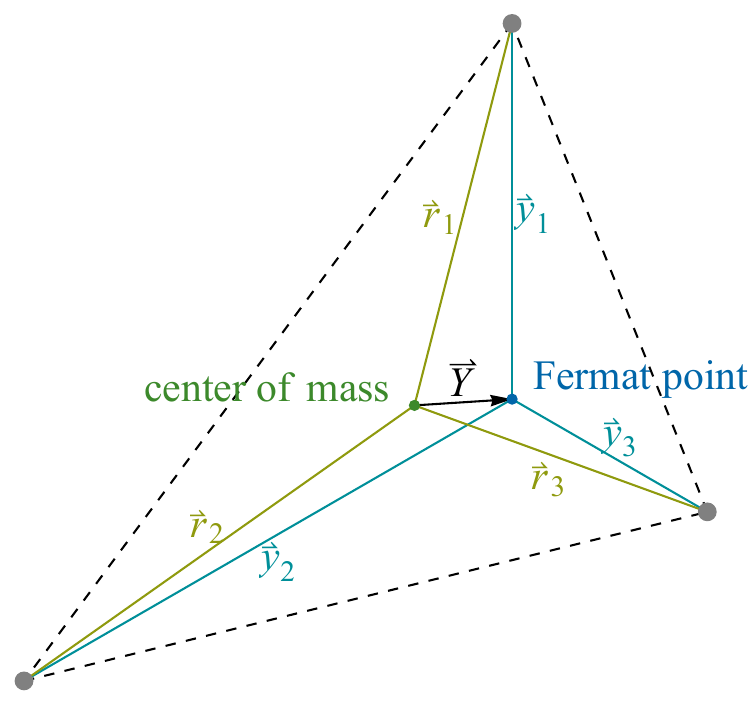}
    \caption{Illustration of Y-shaped (teal solid lines) and $\Delta$-shaped (gray dashed lines) junction configurations. Gray dots indicate the position of bottom quarks, green for the center of mass, and blue for the Fermat point.
    \label{fig:msr}}
\end{figure}
In this section we present the details for computing the mean-squared radius of gauge field energy distribution, assuming either Y- or $\Delta$-shape configurations.
With the settings shown in Fig.~\ref{fig:msr}, we may compute the mean-squared radius of bottom quarks as,
\begin{align}
    \langle r_{\mathrm{bottom}}^2 \rangle = \frac{1}{3}\sum_i r_i^2,
\end{align}
and that of the Y-shape junction as 
\begin{align}
\begin{split}
\langle r_{\mathrm{gauge},\mathrm{Y}}^2 \rangle \equiv\,& 
    \frac{1}{3} \sum_i \int_0^1 |\boldsymbol{Y}+\lambda \boldsymbol{y_i}|^2 \mathrm{d}\lambda \\
=\,&
    \frac{1}{3} \sum_i \int_0^1 |(1-\lambda)\boldsymbol{Y}+\lambda \boldsymbol{r_i}|^2 \mathrm{d}\lambda \\
=\,&
    \frac{1}{3} \sum_i \int_0^1 \Big( (1-\lambda)^2 Y^2 + \lambda^2 {r_i}^2 + 2(\lambda-\lambda^2)\boldsymbol{Y}\cdot \boldsymbol{r_i}\Big)\mathrm{d}\lambda \\
=\,&
    \frac{1}{3} \big(\langle r_{\mathrm{bottom}}^2 \rangle + Y^2\big)\,,
\end{split}
\end{align}
whereas that of the $\Delta$-shape junction is
\begin{align}
\begin{split}
\langle r_{\mathrm{gauge},\Delta}^2 \rangle 
\equiv\,& 
    \frac{1}{6} \sum_{i\neq j} \int_0^1 |(1-\lambda)\boldsymbol{r_i}+\lambda \boldsymbol{r_j}|^2 \mathrm{d}\lambda \\
=\,& 
    \frac{1}{6} \sum_{i \neq j} \frac{r_i^2+r_j^2 + \boldsymbol{r_i}\cdot \boldsymbol{r_j}}{3}\\
=\,& 
    \frac{2}{9} (r_1^2+r_2^2+r_3^2)
    + 
    \frac{\boldsymbol{r_1}\cdot \boldsymbol{r_2} + \boldsymbol{r_2}\cdot \boldsymbol{r_3}+\boldsymbol{r_3}\cdot \boldsymbol{r_1}}{9}\\
=\,& 
    \frac{2}{9} (r_1^2+r_2^2+r_3^2)
    - 
    \frac{r_1^2+r_2^2+r_3^2}{18}\\
=\,&
    \frac{1}{2} \langle r_{\mathrm{bottom}}^2 \rangle\,.
\end{split}
\end{align}

Thus, the $\Delta$-shaped junction predicts a mean-squared radius for the gauge field energy being half of that of the bottom quarks despite of the particular configuration of the bottom quarks; the Y-shaped junction predicts a third for a equilateral-triangle, with extra contributions from the distance-squared between the Fermat point and the center of mass.

\end{widetext}
\end{appendix}

\bibliographystyle{apsrev4-1.bst}
\bibliography{Ref}

\begin{thebibliography}{26}%
\makeatletter
\providecommand \@ifxundefined [1]{%
 \@ifx{#1\undefined}
}%
\providecommand \@ifnum [1]{%
 \ifnum #1\expandafter \@firstoftwo
 \else \expandafter \@secondoftwo
 \fi
}%
\providecommand \@ifx [1]{%
 \ifx #1\expandafter \@firstoftwo
 \else \expandafter \@secondoftwo
 \fi
}%
\providecommand \natexlab [1]{#1}%
\providecommand \enquote  [1]{``#1''}%
\providecommand \bibnamefont  [1]{#1}%
\providecommand \bibfnamefont [1]{#1}%
\providecommand \citenamefont [1]{#1}%
\providecommand \href@noop [0]{\@secondoftwo}%
\providecommand \href [0]{\begingroup \@sanitize@url \@href}%
\providecommand \@href[1]{\@@startlink{#1}\@@href}%
\providecommand \@@href[1]{\endgroup#1\@@endlink}%
\providecommand \@sanitize@url [0]{\catcode `\\12\catcode `\$12\catcode
  `\&12\catcode `\#12\catcode `\^12\catcode `\_12\catcode `\%12\relax}%
\providecommand \@@startlink[1]{}%
\providecommand \@@endlink[0]{}%
\providecommand \url  [0]{\begingroup\@sanitize@url \@url }%
\providecommand \@url [1]{\endgroup\@href {#1}{\urlprefix }}%
\providecommand \urlprefix  [0]{URL }%
\providecommand \Eprint [0]{\href }%
\providecommand \doibase [0]{http://dx.doi.org/}%
\providecommand \selectlanguage [0]{\@gobble}%
\providecommand \bibinfo  [0]{\@secondoftwo}%
\providecommand \bibfield  [0]{\@secondoftwo}%
\providecommand \translation [1]{[#1]}%
\providecommand \BibitemOpen [0]{}%
\providecommand \bibitemStop [0]{}%
\providecommand \bibitemNoStop [0]{.\EOS\space}%
\providecommand \EOS [0]{\spacefactor3000\relax}%
\providecommand \BibitemShut  [1]{\csname bibitem#1\endcsname}%
\let\auto@bib@innerbib\@empty
\bibitem [{\citenamefont {Brambilla}\ \emph {et~al.}(2010)\citenamefont
  {Brambilla}, \citenamefont {Ghiglieri},\ and\ \citenamefont
  {Vairo}}]{Brambilla:2009cd}%
  \BibitemOpen
  \bibfield  {author} {\bibinfo {author} {\bibfnamefont {N.}~\bibnamefont
  {Brambilla}}, \bibinfo {author} {\bibfnamefont {J.}~\bibnamefont
  {Ghiglieri}}, \ and\ \bibinfo {author} {\bibfnamefont {A.}~\bibnamefont
  {Vairo}},\ }\href {\doibase 10.1103/PhysRevD.81.054031} {\bibfield  {journal}
  {\bibinfo  {journal} {Phys. Rev. D}\ }\textbf {\bibinfo {volume} {81}},\
  \bibinfo {pages} {054031} (\bibinfo {year} {2010})},\ \bibinfo {note}
  {[Erratum: Phys.Rev.D 107, 019904 (2023)]},\ \Eprint
  {http://arxiv.org/abs/0911.3541} {arXiv:0911.3541 [hep-ph]} \BibitemShut
  {NoStop}%
\bibitem [{\citenamefont {Brambilla}\ \emph {et~al.}(2005)\citenamefont
  {Brambilla}, \citenamefont {Vairo},\ and\ \citenamefont
  {Rosch}}]{Brambilla:2005yk}%
  \BibitemOpen
  \bibfield  {author} {\bibinfo {author} {\bibfnamefont {N.}~\bibnamefont
  {Brambilla}}, \bibinfo {author} {\bibfnamefont {A.}~\bibnamefont {Vairo}}, \
  and\ \bibinfo {author} {\bibfnamefont {T.}~\bibnamefont {Rosch}},\ }\href
  {\doibase 10.1103/PhysRevD.72.034021} {\bibfield  {journal} {\bibinfo
  {journal} {Phys. Rev. D}\ }\textbf {\bibinfo {volume} {72}},\ \bibinfo
  {pages} {034021} (\bibinfo {year} {2005})},\ \Eprint
  {http://arxiv.org/abs/hep-ph/0506065} {arXiv:hep-ph/0506065} \BibitemShut
  {NoStop}%
\bibitem [{\citenamefont {Silvestre-Brac}(1996)}]{Silvestre-Brac:1996myf}%
  \BibitemOpen
  \bibfield  {author} {\bibinfo {author} {\bibfnamefont {B.}~\bibnamefont
  {Silvestre-Brac}},\ }\href {\doibase 10.1007/s006010050028} {\bibfield
  {journal} {\bibinfo  {journal} {Few Body Syst.}\ }\textbf {\bibinfo {volume}
  {20}},\ \bibinfo {pages} {1} (\bibinfo {year} {1996})}\BibitemShut {NoStop}%
\bibitem [{\citenamefont {Martynenko}(2008)}]{Martynenko:2007je}%
  \BibitemOpen
  \bibfield  {author} {\bibinfo {author} {\bibfnamefont {A.~P.}\ \bibnamefont
  {Martynenko}},\ }\href {\doibase 10.1016/j.physletb.2008.04.030} {\bibfield
  {journal} {\bibinfo  {journal} {Phys. Lett. B}\ }\textbf {\bibinfo {volume}
  {663}},\ \bibinfo {pages} {317} (\bibinfo {year} {2008})},\ \Eprint
  {http://arxiv.org/abs/0708.2033} {arXiv:0708.2033 [hep-ph]} \BibitemShut
  {NoStop}%
\bibitem [{\citenamefont {Vijande}\ \emph {et~al.}(2015)\citenamefont
  {Vijande}, \citenamefont {Valcarce},\ and\ \citenamefont
  {Garcilazo}}]{Vijande:2015faa}%
  \BibitemOpen
  \bibfield  {author} {\bibinfo {author} {\bibfnamefont {J.}~\bibnamefont
  {Vijande}}, \bibinfo {author} {\bibfnamefont {A.}~\bibnamefont {Valcarce}}, \
  and\ \bibinfo {author} {\bibfnamefont {H.}~\bibnamefont {Garcilazo}},\ }\href
  {\doibase 10.1103/PhysRevD.91.054011} {\bibfield  {journal} {\bibinfo
  {journal} {Phys. Rev. D}\ }\textbf {\bibinfo {volume} {91}},\ \bibinfo
  {pages} {054011} (\bibinfo {year} {2015})},\ \Eprint
  {http://arxiv.org/abs/1507.03735} {arXiv:1507.03735 [hep-ph]} \BibitemShut
  {NoStop}%
\bibitem [{\citenamefont {Takahashi}\ \emph {et~al.}(2001)\citenamefont
  {Takahashi}, \citenamefont {Matsufuru}, \citenamefont {Nemoto},\ and\
  \citenamefont {Suganuma}}]{Takahashi:2000te}%
  \BibitemOpen
  \bibfield  {author} {\bibinfo {author} {\bibfnamefont {T.~T.}\ \bibnamefont
  {Takahashi}}, \bibinfo {author} {\bibfnamefont {H.}~\bibnamefont
  {Matsufuru}}, \bibinfo {author} {\bibfnamefont {Y.}~\bibnamefont {Nemoto}}, \
  and\ \bibinfo {author} {\bibfnamefont {H.}~\bibnamefont {Suganuma}},\ }\href
  {\doibase 10.1103/PhysRevLett.86.18} {\bibfield  {journal} {\bibinfo
  {journal} {Phys. Rev. Lett.}\ }\textbf {\bibinfo {volume} {86}},\ \bibinfo
  {pages} {18} (\bibinfo {year} {2001})},\ \Eprint
  {http://arxiv.org/abs/hep-lat/0006005} {arXiv:hep-lat/0006005} \BibitemShut
  {NoStop}%
\bibitem [{\citenamefont {Takahashi}\ \emph {et~al.}(2002)\citenamefont
  {Takahashi}, \citenamefont {Suganuma}, \citenamefont {Nemoto},\ and\
  \citenamefont {Matsufuru}}]{Takahashi:2002bw}%
  \BibitemOpen
  \bibfield  {author} {\bibinfo {author} {\bibfnamefont {T.~T.}\ \bibnamefont
  {Takahashi}}, \bibinfo {author} {\bibfnamefont {H.}~\bibnamefont {Suganuma}},
  \bibinfo {author} {\bibfnamefont {Y.}~\bibnamefont {Nemoto}}, \ and\ \bibinfo
  {author} {\bibfnamefont {H.}~\bibnamefont {Matsufuru}},\ }\href {\doibase
  10.1103/PhysRevD.65.114509} {\bibfield  {journal} {\bibinfo  {journal} {Phys.
  Rev. D}\ }\textbf {\bibinfo {volume} {65}},\ \bibinfo {pages} {114509}
  (\bibinfo {year} {2002})},\ \Eprint {http://arxiv.org/abs/hep-lat/0204011}
  {arXiv:hep-lat/0204011} \BibitemShut {NoStop}%
\bibitem [{\citenamefont {Meinel}(2012)}]{Meinel:2012qz}%
  \BibitemOpen
  \bibfield  {author} {\bibinfo {author} {\bibfnamefont {S.}~\bibnamefont
  {Meinel}},\ }\href {\doibase 10.1103/PhysRevD.85.114510} {\bibfield
  {journal} {\bibinfo  {journal} {Phys. Rev. D}\ }\textbf {\bibinfo {volume}
  {85}},\ \bibinfo {pages} {114510} (\bibinfo {year} {2012})},\ \Eprint
  {http://arxiv.org/abs/1202.1312} {arXiv:1202.1312 [hep-lat]} \BibitemShut
  {NoStop}%
\bibitem [{\citenamefont {Caswell}\ and\ \citenamefont
  {Lepage}(1986)}]{Caswell:1985ui}%
  \BibitemOpen
  \bibfield  {author} {\bibinfo {author} {\bibfnamefont {W.~E.}\ \bibnamefont
  {Caswell}}\ and\ \bibinfo {author} {\bibfnamefont {G.~P.}\ \bibnamefont
  {Lepage}},\ }\href {\doibase 10.1016/0370-2693(86)91297-9} {\bibfield
  {journal} {\bibinfo  {journal} {Phys. Lett. B}\ }\textbf {\bibinfo {volume}
  {167}},\ \bibinfo {pages} {437} (\bibinfo {year} {1986})}\BibitemShut
  {NoStop}%
\bibitem [{\citenamefont {Brambilla}\ \emph {et~al.}(2000)\citenamefont
  {Brambilla}, \citenamefont {Pineda}, \citenamefont {Soto},\ and\
  \citenamefont {Vairo}}]{Brambilla:1999xf}%
  \BibitemOpen
  \bibfield  {author} {\bibinfo {author} {\bibfnamefont {N.}~\bibnamefont
  {Brambilla}}, \bibinfo {author} {\bibfnamefont {A.}~\bibnamefont {Pineda}},
  \bibinfo {author} {\bibfnamefont {J.}~\bibnamefont {Soto}}, \ and\ \bibinfo
  {author} {\bibfnamefont {A.}~\bibnamefont {Vairo}},\ }\href {\doibase
  10.1016/S0550-3213(99)00693-8} {\bibfield  {journal} {\bibinfo  {journal}
  {Nucl. Phys. B}\ }\textbf {\bibinfo {volume} {566}},\ \bibinfo {pages} {275}
  (\bibinfo {year} {2000})},\ \Eprint {http://arxiv.org/abs/hep-ph/9907240}
  {arXiv:hep-ph/9907240} \BibitemShut {NoStop}%
\bibitem [{\citenamefont {Brambilla}\ \emph {et~al.}(2013)\citenamefont
  {Brambilla}, \citenamefont {Karbstein},\ and\ \citenamefont
  {Vairo}}]{Brambilla:2013vx}%
  \BibitemOpen
  \bibfield  {author} {\bibinfo {author} {\bibfnamefont {N.}~\bibnamefont
  {Brambilla}}, \bibinfo {author} {\bibfnamefont {F.}~\bibnamefont
  {Karbstein}}, \ and\ \bibinfo {author} {\bibfnamefont {A.}~\bibnamefont
  {Vairo}},\ }\href {\doibase 10.1103/PhysRevD.87.074014} {\bibfield  {journal}
  {\bibinfo  {journal} {Phys. Rev. D}\ }\textbf {\bibinfo {volume} {87}},\
  \bibinfo {pages} {074014} (\bibinfo {year} {2013})},\ \Eprint
  {http://arxiv.org/abs/1301.3013} {arXiv:1301.3013 [hep-ph]} \BibitemShut
  {NoStop}%
\bibitem [{\citenamefont {Zhao}\ \emph {et~al.}(2020)\citenamefont {Zhao},
  \citenamefont {Zhou}, \citenamefont {Chen},\ and\ \citenamefont
  {Zhuang}}]{Zhao:2020jqu}%
  \BibitemOpen
  \bibfield  {author} {\bibinfo {author} {\bibfnamefont {J.}~\bibnamefont
  {Zhao}}, \bibinfo {author} {\bibfnamefont {K.}~\bibnamefont {Zhou}}, \bibinfo
  {author} {\bibfnamefont {S.}~\bibnamefont {Chen}}, \ and\ \bibinfo {author}
  {\bibfnamefont {P.}~\bibnamefont {Zhuang}},\ }\href {\doibase
  10.1016/j.ppnp.2020.103801} {\bibfield  {journal} {\bibinfo  {journal} {Prog.
  Part. Nucl. Phys.}\ }\textbf {\bibinfo {volume} {114}},\ \bibinfo {pages}
  {103801} (\bibinfo {year} {2020})},\ \Eprint
  {http://arxiv.org/abs/2005.08277} {arXiv:2005.08277 [nucl-th]} \BibitemShut
  {NoStop}%
\bibitem [{\citenamefont {Liu}\ \emph {et~al.}(2020)\citenamefont {Liu},
  \citenamefont {L\"u},\ and\ \citenamefont {Zhong}}]{Liu:2019vtx}%
  \BibitemOpen
  \bibfield  {author} {\bibinfo {author} {\bibfnamefont {M.-S.}\ \bibnamefont
  {Liu}}, \bibinfo {author} {\bibfnamefont {Q.-F.}\ \bibnamefont {L\"u}}, \
  and\ \bibinfo {author} {\bibfnamefont {X.-H.}\ \bibnamefont {Zhong}},\ }\href
  {\doibase 10.1103/PhysRevD.101.074031} {\bibfield  {journal} {\bibinfo
  {journal} {Phys. Rev. D}\ }\textbf {\bibinfo {volume} {101}},\ \bibinfo
  {pages} {074031} (\bibinfo {year} {2020})},\ \Eprint
  {http://arxiv.org/abs/1912.11805} {arXiv:1912.11805 [hep-ph]} \BibitemShut
  {NoStop}%
\bibitem [{\citenamefont {Yang}\ \emph {et~al.}(2020)\citenamefont {Yang},
  \citenamefont {Ping}, \citenamefont {Ortega},\ and\ \citenamefont
  {Segovia}}]{Yang:2019lsg}%
  \BibitemOpen
  \bibfield  {author} {\bibinfo {author} {\bibfnamefont {G.}~\bibnamefont
  {Yang}}, \bibinfo {author} {\bibfnamefont {J.}~\bibnamefont {Ping}}, \bibinfo
  {author} {\bibfnamefont {P.~G.}\ \bibnamefont {Ortega}}, \ and\ \bibinfo
  {author} {\bibfnamefont {J.}~\bibnamefont {Segovia}},\ }\href {\doibase
  10.1088/1674-1137/44/2/023102} {\bibfield  {journal} {\bibinfo  {journal}
  {Chin. Phys. C}\ }\textbf {\bibinfo {volume} {44}},\ \bibinfo {pages}
  {023102} (\bibinfo {year} {2020})},\ \Eprint
  {http://arxiv.org/abs/1904.10166} {arXiv:1904.10166 [hep-ph]} \BibitemShut
  {NoStop}%
\bibitem [{\citenamefont {Shi}\ \emph {et~al.}(2020)\citenamefont {Shi},
  \citenamefont {Zhao},\ and\ \citenamefont {Zhuang}}]{Shi:2019tji}%
  \BibitemOpen
  \bibfield  {author} {\bibinfo {author} {\bibfnamefont {S.}~\bibnamefont
  {Shi}}, \bibinfo {author} {\bibfnamefont {J.}~\bibnamefont {Zhao}}, \ and\
  \bibinfo {author} {\bibfnamefont {P.}~\bibnamefont {Zhuang}},\ }\href
  {\doibase 10.1088/1674-1137/44/8/084101} {\bibfield  {journal} {\bibinfo
  {journal} {Chin. Phys. C}\ }\textbf {\bibinfo {volume} {44}},\ \bibinfo
  {pages} {084101} (\bibinfo {year} {2020})},\ \Eprint
  {http://arxiv.org/abs/1905.10627} {arXiv:1905.10627 [nucl-th]} \BibitemShut
  {NoStop}%
\bibitem [{\citenamefont {Kawanai}\ and\ \citenamefont
  {Sasaki}(2012)}]{Kawanai:2011jt}%
  \BibitemOpen
  \bibfield  {author} {\bibinfo {author} {\bibfnamefont {T.}~\bibnamefont
  {Kawanai}}\ and\ \bibinfo {author} {\bibfnamefont {S.}~\bibnamefont
  {Sasaki}},\ }\href {\doibase 10.1103/PhysRevD.85.091503} {\bibfield
  {journal} {\bibinfo  {journal} {Phys. Rev. D}\ }\textbf {\bibinfo {volume}
  {85}},\ \bibinfo {pages} {091503} (\bibinfo {year} {2012})},\ \Eprint
  {http://arxiv.org/abs/1110.0888} {arXiv:1110.0888 [hep-lat]} \BibitemShut
  {NoStop}%
\bibitem [{\citenamefont {Takahashi}\ and\ \citenamefont
  {Suganuma}(2004)}]{Takahashi:2004rw}%
  \BibitemOpen
  \bibfield  {author} {\bibinfo {author} {\bibfnamefont {T.~T.}\ \bibnamefont
  {Takahashi}}\ and\ \bibinfo {author} {\bibfnamefont {H.}~\bibnamefont
  {Suganuma}},\ }\href {\doibase 10.1103/PhysRevD.70.074506} {\bibfield
  {journal} {\bibinfo  {journal} {Phys. Rev. D}\ }\textbf {\bibinfo {volume}
  {70}},\ \bibinfo {pages} {074506} (\bibinfo {year} {2004})},\ \Eprint
  {http://arxiv.org/abs/hep-lat/0409105} {arXiv:hep-lat/0409105} \BibitemShut
  {NoStop}%
\bibitem [{\citenamefont {Suganuma}\ \emph {et~al.}(2000)\citenamefont
  {Suganuma}, \citenamefont {Takahashi}, \citenamefont {Matsufuru},\ and\
  \citenamefont {Nemoto}}]{Suganuma:2000bi}%
  \BibitemOpen
  \bibfield  {author} {\bibinfo {author} {\bibfnamefont {H.}~\bibnamefont
  {Suganuma}}, \bibinfo {author} {\bibfnamefont {T.~T.}\ \bibnamefont
  {Takahashi}}, \bibinfo {author} {\bibfnamefont {H.}~\bibnamefont
  {Matsufuru}}, \ and\ \bibinfo {author} {\bibfnamefont {Y.}~\bibnamefont
  {Nemoto}},\ }\href {\doibase 10.1016/S0375-9474(00)00408-5} {\bibfield
  {journal} {\bibinfo  {journal} {Nucl. Phys. A}\ }\textbf {\bibinfo {volume}
  {680}},\ \bibinfo {pages} {159} (\bibinfo {year} {2000})},\ \Eprint
  {http://arxiv.org/abs/hep-lat/0205029} {arXiv:hep-lat/0205029} \BibitemShut
  {NoStop}%
\bibitem [{\citenamefont {Zhao}\ and\ \citenamefont
  {Shi}(2024{\natexlab{a}})}]{Zhao:2023qww}%
  \BibitemOpen
  \bibfield  {author} {\bibinfo {author} {\bibfnamefont {J.}~\bibnamefont
  {Zhao}}\ and\ \bibinfo {author} {\bibfnamefont {S.}~\bibnamefont {Shi}},\
  }\href {\doibase 10.1103/PhysRevC.109.024901} {\bibfield  {journal} {\bibinfo
   {journal} {Phys. Rev. C}\ }\textbf {\bibinfo {volume} {109}},\ \bibinfo
  {pages} {024901} (\bibinfo {year} {2024}{\natexlab{a}})},\ \Eprint
  {http://arxiv.org/abs/2311.04594} {arXiv:2311.04594 [nucl-th]} \BibitemShut
  {NoStop}%
\bibitem [{\citenamefont {Crater}(1994)}]{CRATER1994470}%
  \BibitemOpen
  \bibfield  {author} {\bibinfo {author} {\bibfnamefont {H.}~\bibnamefont
  {Crater}},\ }\href {\doibase https://doi.org/10.1006/jcph.1994.1211}
  {\bibfield  {journal} {\bibinfo  {journal} {Journal of Computational
  Physics}\ }\textbf {\bibinfo {volume} {115}},\ \bibinfo {pages} {470}
  (\bibinfo {year} {1994})}\BibitemShut {NoStop}%
\bibitem [{\citenamefont {Zhao}\ and\ \citenamefont
  {Shi}(2024{\natexlab{b}})}]{Zhao:2024atb}%
  \BibitemOpen
  \bibfield  {author} {\bibinfo {author} {\bibfnamefont {J.}~\bibnamefont
  {Zhao}}\ and\ \bibinfo {author} {\bibfnamefont {S.}~\bibnamefont {Shi}},\
  }\href {\doibase 10.1016/j.cpc.2024.109284} {\bibfield  {journal} {\bibinfo
  {journal} {Comput. Phys. Commun.}\ }\textbf {\bibinfo {volume} {303}},\
  \bibinfo {pages} {109284} (\bibinfo {year} {2024}{\natexlab{b}})},\ \Eprint
  {http://arxiv.org/abs/2403.02747} {arXiv:2403.02747 [physics.comp-ph]}
  \BibitemShut {NoStop}%
\bibitem [{\citenamefont {Raynal}\ and\ \citenamefont
  {Revai}(1970)}]{Raynal:1970ah}%
  \BibitemOpen
  \bibfield  {author} {\bibinfo {author} {\bibfnamefont {J.}~\bibnamefont
  {Raynal}}\ and\ \bibinfo {author} {\bibfnamefont {J.}~\bibnamefont {Revai}},\
  }\href {\doibase 10.1007/BF02756127} {\bibfield  {journal} {\bibinfo
  {journal} {Nuovo Cim. A}\ }\textbf {\bibinfo {volume} {68}},\ \bibinfo
  {pages} {612} (\bibinfo {year} {1970})}\BibitemShut {NoStop}%
\bibitem [{\citenamefont {Zyla}\ \emph {et~al.}(2020)\citenamefont {Zyla} \emph
  {et~al.}}]{10.1093/ptep/ptaa104}%
  \BibitemOpen
  \bibfield  {author} {\bibinfo {author} {\bibfnamefont {P.~A.}\ \bibnamefont
  {Zyla}} \emph {et~al.},\ }\href {\doibase 10.1093/ptep/ptaa104} {\bibfield
  {journal} {\bibinfo  {journal} {Progress of Theoretical and Experimental
  Physics}\ }\textbf {\bibinfo {volume} {2020}},\ \bibinfo {pages} {083C01}
  (\bibinfo {year} {2020})},\ \Eprint
  {http://arxiv.org/abs/https://academic.oup.com/ptep/article-pdf/2020/8/083C01/34673722/ptaa104.pdf}
  {https://academic.oup.com/ptep/article-pdf/2020/8/083C01/34673722/ptaa104.pdf}
  \BibitemShut {NoStop}%
\bibitem [{\citenamefont {Ashman}\ \emph {et~al.}(1988)\citenamefont {Ashman}
  \emph {et~al.}}]{EuropeanMuon:1987isl}%
  \BibitemOpen
  \bibfield  {author} {\bibinfo {author} {\bibfnamefont {J.}~\bibnamefont
  {Ashman}} \emph {et~al.} (\bibinfo {collaboration} {European Muon}),\ }\href
  {\doibase 10.1016/0370-2693(88)91523-7} {\bibfield  {journal} {\bibinfo
  {journal} {Phys. Lett. B}\ }\textbf {\bibinfo {volume} {206}},\ \bibinfo
  {pages} {364} (\bibinfo {year} {1988})}\BibitemShut {NoStop}%
\bibitem [{\citenamefont {Aboona}\ \emph {et~al.}(2026)\citenamefont {Aboona}
  \emph {et~al.}}]{STAR:2024lvy}%
  \BibitemOpen
  \bibfield  {author} {\bibinfo {author} {\bibfnamefont {B.~E.}\ \bibnamefont
  {Aboona}} \emph {et~al.} (\bibinfo {collaboration} {STAR}),\ }\href {\doibase
  10.1126/science.ads5962} {\bibfield  {journal} {\bibinfo  {journal}
  {Science}\ }\textbf {\bibinfo {volume} {393}},\ \bibinfo {pages} {ads5962}
  (\bibinfo {year} {2026})},\ \Eprint {http://arxiv.org/abs/2408.15441}
  {arXiv:2408.15441 [nucl-ex]} \BibitemShut {NoStop}%
\bibitem [{\citenamefont {Li}\ \emph {et~al.}(2026)\citenamefont {Li},
  \citenamefont {Shi},\ and\ \citenamefont {Zhuang}}]{LiXiaQing:2026}%
  \BibitemOpen
  \bibfield  {author} {\bibinfo {author} {\bibfnamefont {X.-Q.}\ \bibnamefont
  {Li}}, \bibinfo {author} {\bibfnamefont {S.-Z.}\ \bibnamefont {Shi}}, \ and\
  \bibinfo {author} {\bibfnamefont {P.-F.}\ \bibnamefont {Zhuang}},\ }\href
  {\doibase 10.1007/s41365-026-02050-w} {\bibfield  {journal} {\bibinfo
  {journal} {Nuclear Science and Techniques}\ }\textbf {\bibinfo {volume}
  {37}},\ \bibinfo {pages} {241} (\bibinfo {year} {2026})}\BibitemShut
  {NoStop}%
\end{thebibliography}%

\end{document}